\documentclass[iop,numberedappendix]{emulateapj}
\usepackage{natbib}
\usepackage{graphicx}
\usepackage{url}
\usepackage[usenames,dvipsnames]{xcolor}
\usepackage{hyperref}
\hypersetup{
    colorlinks=true, 
    linktoc=all,     
    linkcolor={blue!60!black}, citecolor={blue!60!black}, urlcolor={blue!60!black} 
}

\newcommand{\Itsz}{\mbox{$I_{\mbox{\tiny tSZE}}$}}
\newcommand{\Iksz}{\mbox{$I_{\mbox{\tiny kSZE}}$}}
\newcommand{\sigT}{\mbox{$\sigma_{\mbox{\tiny T}}$}}
\newcommand{\Tcmb}{\mbox{$T_{\mbox{\tiny CMB}}$}}
\newcommand{\kB}{\mbox{$k_{\mbox{\tiny B}}$}}

\newcommand{\Lamee}{\mbox{$\Lambda_{\mbox{\tiny ee}}$}}

\newcommand{\Pe}{\mbox{$P_{\mbox{\scriptsize e}}$}}
\newcommand{\etal}{{\it et al.}}

\newcommand{\uJybm}{$\, \mu$Jy beam$^{-1}$}

\newcommand{\chandra}{{\it Chandra}}

\newcommand{\macsc}{MACS J0717.5+3745}

\newcommand{\Tx}{\mbox{$T_{\mbox{\tiny X}}$}}
\newcommand{\te}{\mbox{$T_{\mbox{\tiny e}}$}}
\newcommand{\mec}{\mbox{$m_{\mbox{\tiny e}} c^2$}}
\newcommand{\dene}{\mbox{$n_{\mbox{\tiny e}}$}}
\newcommand{\denesq}{\mbox{$n^2_{\mbox{\tiny e}}$}}

\newcommand{\ysze}{\mbox{$Y_{\mbox{\tiny tSZE}}$}}
\newcommand{\sx}{\mbox{$S_{\mbox{\tiny X}}$}}

\slugcomment{}
\shortauthors{Mroczkowski \etal}
\shorttitle{Multi-wavelength SZE imaging of MACS~J0717.5+3745}
\begin{document}
\title{A Multi-wavelength Study of the Sunyaev--Zel'dovich Effect 
in the Triple-Merger Cluster MACS~J0717.5+3745 with MUSTANG and Bolocam}
\author{
  Tony Mroczkowski\altaffilmark{1,2,9},
  Simon Dicker\altaffilmark{3},
  Jack Sayers\altaffilmark{2},
  Erik D.\ Reese\altaffilmark{3},
  Brian Mason\altaffilmark{4},
  Nicole Czakon\altaffilmark{2},
  Charles Romero\altaffilmark{4,5},
  Alexander Young\altaffilmark{3},
  Mark Devlin\altaffilmark{3},
  Sunil Golwala\altaffilmark{2},
  Phillip Korngut\altaffilmark{1,2},
  Craig Sarazin\altaffilmark{5},\\
  James Bock\altaffilmark{1,2},
  Patrick M.\ Koch\altaffilmark{6},
  Kai-Yang Lin\altaffilmark{6},
  Sandor M.\ Molnar\altaffilmark{7},
  Elena Pierpaoli\altaffilmark{8},
  Keiichi Umetsu\altaffilmark{6},
  and
  Michael Zemcov\altaffilmark{1,2}
} 
\altaffiltext{1}{Jet Propulsion Laboratory, 4800 Oak Grove Drive, 
Pasadena, CA 91109, USA; \email{tonym@astro.caltech.edu}}
\altaffiltext{2}{California Institute of Technology, 1200 East California Boulevard, 
Pasadena, CA 91125, USA}
\altaffiltext{3}{Department of Physics and Astronomy, University of
  Pennsylvania, 209 South 33rd Street, Philadelphia, PA, 19104, USA}
\altaffiltext{4}{National Radio Astronomy Observatory, 520 Edgemont Road, 
Charlottesville, VA 22903, USA}
\altaffiltext{5}{Department of Astronomy, University of Virginia, 
P.O. Box 400325, Charlottesville, VA 22901, USA}
\altaffiltext{6}{Institute of Astronomy and Astrophysics, Academia Sinica,
P. O. Box 23-141, Taipei 10617, Taiwan}
\altaffiltext{7}{Leung Center for Cosmology and Particle Astrophysics, 
National Taiwan University, Taipei 10617, Taiwan, Republic of China}
\altaffiltext{8}{Department of Physics and Astronomy, University of Southern 
California, Los Angeles, CA 90089, USA}
{\altaffiltext{9}{NASA Einstein Postdoctoral Fellow}
\begin{abstract}

We present 90, 140, and 268~GHz sub-arcminute resolution imaging
of the Sunyaev-Zel'dovich effect (SZE) in the disturbed, intermediate redshift 
($z=0.5458$) galaxy cluster \macsc, a triple-merger system comprising four 
distinct, optically-detected subclusters.
Our 90~GHz SZE data result in a sensitive, 34~\uJybm\ map of the SZE at 
13$^{\prime\prime}$ effective resolution using the MUSTANG bolometer array on 
the Green Bank Telescope (GBT).
Our 140 and 268~GHz SZE imaging, with resolutions of 58$^{\prime\prime}$ and 
31$^{\prime\prime}$ and sensitivities of 1.8 and 3.3~mJy beam$^{-1}$,
respectively, was obtained through observations from the 
Caltech Submillimeter Observatory using Bolocam.
We compare these maps to a two-dimensional pressure map derived from \chandra\
X-ray observations.
Our MUSTANG SZE data confirm previous indications from \chandra\ of a pressure 
enhancement due to shock-heated, $\gtrsim 20 \, {\rm keV}$ gas immediately 
adjacent to extended radio emission seen in low-frequency radio maps of this cluster. 
MUSTANG also detects pressure substructure that is not well-constrained 
by the \chandra\ X-ray data in the remnant core of a merging subcluster. 
We find that the small-scale pressure enhancements in the MUSTANG data 
amount to $\sim 2$\% of the total pressure measured in the 140~GHz Bolocam 
observations.
The X-ray inferred pseudo-pressure template also fails on larger scales to 
accurately describe the Bolocam data, particularly at the location of the 
subcluster with a remnant core known to have a high line of 
sight optical velocity of $\sim 3200$~km s$^{-1}$.  
Our Bolocam data are adequately described when we add an additional component---not 
described by a thermal SZE spectrum---to the X-ray template coincident 
with this subcluster.
Using flux densities extracted from our model fits, and marginalizing over the X-ray 
spectroscopic temperature constraints for the region, we fit a thermal + kinetic 
SZE spectrum to our Bolocam data and find that the subcluster has a 
best-fit line-of-sight proper velocity $v_z = 3600^{+3440}_{-2160}$~km s$^{-1}$,
in agreement with the optical velocity estimates for the subcluster.  
The probability $v_z \leq 0$ given our measurements is 2.1\%.
Repeating this analysis using flux densities measured directly from our maps 
results in a 3.4\% probability $v_z \leq 0$.
We note that this tantalizing result for the kinetic SZE is on resolved, subcluster
scales.
\end{abstract}
\section{Introduction}
\label{sec:intro}
\setcounter{footnote}{0}
Massive ($\gtrsim 10^{15}~M_\odot$) galaxy clusters are the 
largest gravitationally-bound objects in the universe 
and their formation is thought to be driven
by mergers of smaller clusters in what are the most energetic events
to take place since the big bang \citep{sarazin2005}.
Our understanding of the astrophysics of mergers and the intracluster medium 
(ICM) has traditionally been advanced through X-ray observations
\citep[e.g.,][]{mcnamara2005,markevitch2007}, which are sensitive
to both the density and the temperature of the gas.
However, sensitive, high angular resolution imaging of the Sunyaev-Zel'dovich 
effect (SZE) has recently become possible \citep[e.g.,][]{kitayama2004,nord2009,
mason2010,korngut2011,plagge2012}, helping to yield a more complete view of the
complex processes in the ICM, particularly at high redshift.

The SZE is due to inverse Compton scattering of cosmic microwave
background (CMB) photons to higher energies on average by hot electrons in 
the ICM \citep{sunyaev1972}; 
for reviews of the SZE, including its thermal and kinetic 
components, see \cite{birkinshaw1999} and \cite{carlstrom2002}. 
The SZE complements X-ray and optical observations, offering some
of its own unique advantages.  
First, unlike intrinsic emission mechanisms (e.g., X-ray, optical, and radio)  
from a cluster, the SZE does not suffer cosmological surface 
brightness dimming ($\propto (1+z)^{-4}$).  
Second, the thermal SZE (tSZE) is proportional to the line-of-sight 
integrated thermal electron pressure, while X-ray observations are 
sensitive to the ICM density-squared.
The different line-of-sight dependences of X-ray and SZE data allow one to 
infer information about the line-of-sight properties of the ICM. 
Third, there is the potential to constrain the 
component of ICM proper velocity along the line-of-sight using
a Doppler shift of the CMB known as the kinetic SZE (kSZE).
The combination of the SZE observations with those from radio, optical,
X-ray, and other wave bands can provide a more complete understanding 
of cluster mergers and other astrophysical processes.

A particularly striking example of a cluster merger is
\macsc\ \citep{ebeling2001,edge2003}. At a redshift $z=0.5458$,
\macsc\ is the hottest member of the MAssive Cluster Survey 
\citep[MACS; ][]{ebeling2001} at $z>0.5$;  \cite{ebeling2007} report an average X-ray 
spectroscopic temperature $\kB \Tx = 11.6 \pm 0.5 \, {\rm  keV}$, 
determined within an overdensity of 1000 times the critical
density of the universe and by excluding the central 70~kpc of the cluster,
and a velocity dispersion $\sigma \sim 1660$ km s$^{-1}$ within a 1~Mpc aperture.
Optical imaging and spectroscopy have shown that
the cluster is located at the end of a $4 \, h_{70}^{-1} \, {\rm Mpc}$ long
filamentary overdensity of galaxies \citep{ebeling2004}, consistent with
the expectation that clusters form at intersections in the cosmic
web. \cite{ma2008,ma2009} have also shown that the cluster
itself comprises four distinct groups of galaxies and appears to be a rare
triple-merger in progress. Low-frequency radio images
\citep{edge2003,vanweeren2009,bonafede2009} show the presence of diffuse radio
emission as well, indicative of a radio relic or halo.

In this paper we compare multi-wavelength observations of the 
SZE with X-ray, radio, and lensing observations. 
Our SZE imaging data include 90~GHz data collected with the MUSTANG bolometer
array \citep{dicker2008} on the Green Bank Telescope (GBT), along with 140 and 
268~GHz data collected with Bolocam \citep{haig2004} from the Caltech 
Submillimeter Observatory (CSO). 
The MUSTANG data have a resolution of $13^{\prime\prime}$---due to the 
$8^{\prime\prime}.5$ instrument resolution and $10^{\prime\prime}$ smoothing of 
the maps---and a sensitivity of $\sim 34$~\uJybm\ within a central radius of 
$\sim 40^{\prime\prime}$. 
The Bolocam data have resolutions of $58^{\prime\prime}$ and $31^{\prime\prime}$ 
at 140 and 268~GHz---on opposite sides of the null in the tSZE spectrum---and
have sensitivities of 1.8 and 3.3~mJy beam$^{-1}$, respectively. 
The Bolocam data probe scales as large as 14$^\prime$, while
the more sensitive MUSTANG data do not reliably probe scales $>1^\prime$.
Additionally, we compare these observations to a CARMA/SZA 31~GHz observation
of this cluster.  The short baselines of CARMA/SZA probe scales up to 12$^\prime$
with a 2$^\prime$ synthesized beam.  
The longer baselines of the CARMA/SZA, with a $\sim 10^{\prime\prime}$ synthesized beam, 
lack the sensitivity to detect SZE on these scales, but provide constraints on 
the locations and spectral indices of the potentially 
contaminating compact radio source population.
We therefore use the CARMA/SZA data to place constraints on the large angular 
scale cluster properties (the ``bulk'' SZE flux) and to help constrain
contamination by radio sources.

The organization of the paper is as follows.  In Section \ref{sec:0717}, we
provide an overview of results from previous optical, low frequency radio,
and X-ray observations of \macsc. In Section \ref{sec:obs}, we describe the MUSTANG,
Bolocam, CARMA/SZA, and \chandra\ X-ray observations and data reduction.
Section \ref{sec:analysis} gives the details of modelling of the 
SZE observable properties and the X-ray data products used in this analysis.
In Section \ref{sec:ksze}, we use the Bolocam 140 and 268 GHz data, along with our X-ray 
temperature constraints, to infer the peculiar velocity of one subcluster 
component with a spectrum that is not well described by the thermal SZE 
alone, and we compare these estimates to those for the most massive subcluster
in this system.
We present our conclusions from this multi-wavelength SZE study in 
Section \ref{sec:conc}.   
Throughout this paper, we adopt a flat, $\Lambda$-dominated cosmology with
$\Omega_{\mbox{\tiny \rm  M}} = 0.3$, $\Omega_\Lambda = 0.7$, 
and $H_0 = 70$ km s$^{-1}$ Mpc$^{-1}$ consistent with recent 
{\it Wilkinson Microwave Anisotropy Probe (WMAP)}  results 
\citep{komatsu2009, komatsu2011}.

\section{Previous Analyses of \macsc}
\label{sec:0717}
 
\macsc, a complex merging system discovered in the MACS survey 
\citep{ebeling2001,ebeling2007}, is among the best-studied massive clusters at 
redshift $z>0.5$, with observations spanning a broad range of the 
electromagnetic spectrum.  In support of our multi-wavelength SZE study, 
we compare our MUSTANG and Bolocam observations to X-ray, optical, and low 
frequency radio data, and to the results of many studies that have relied
upon these data.

\begin{figure}
  \centerline{
    \includegraphics[width=3.3in]{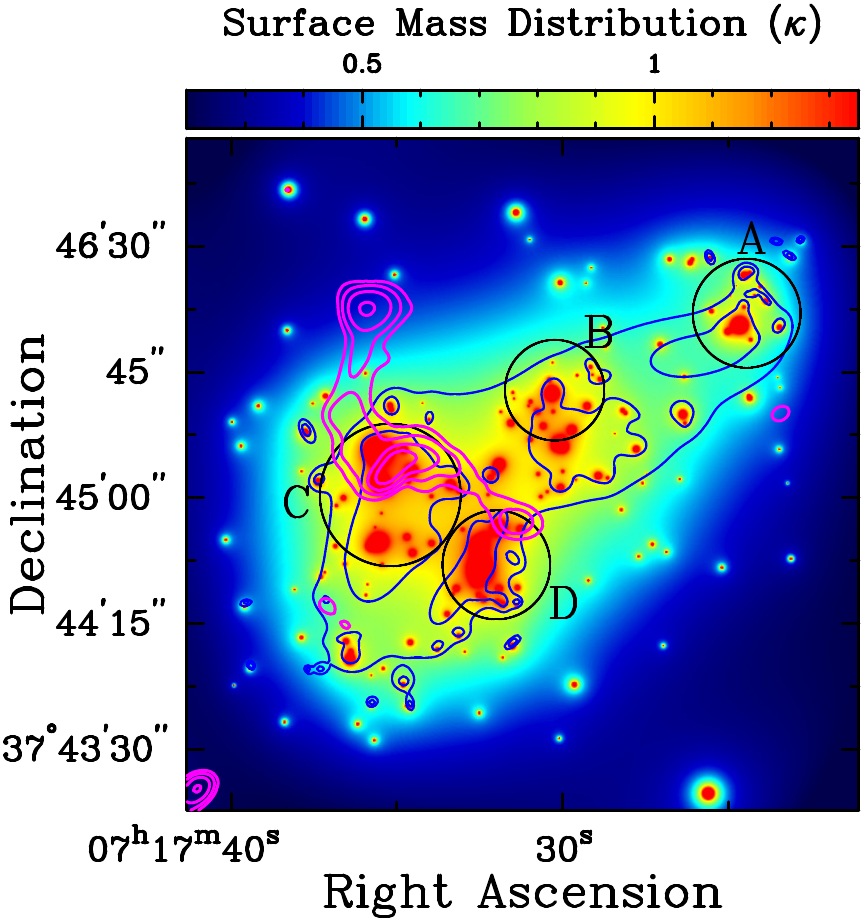}
  }
  \caption{Surface mass distribution in units of the critical surface density 
    (convergence, $\kappa$) for \macsc\ determined by 
    \cite{zitrin2009} through strong lensing measurements (color image).  
    Black circles correspond to the four main peaks in the cluster light 
    distribution as identified by \cite{ma2009}. 
    The magenta contours (3$\sigma$, 6$\sigma$, 12$\sigma$, 24$\sigma$) 
    reproduce the 610~MHz GMRT data in \cite{vanweeren2009}.
    The blue contours reproduce the (5,7)$\times10^{10}~M_\odot \,\rm arcsec^{-2}$
    contours in \cite{limousin2012}.
    We note that recent redshift information reported in \cite{limousin2012} 
    reduces the overall mass estimate, but that the features otherwise show 
    broad agreement.
    \label{fig:lensing}}
\end{figure}

From X-ray and optical analyses, \cite{ma2009} identify four distinct components
in \macsc.
These subclusters are shown in Figure~\ref{fig:lensing} 
on the strong lensing data from \cite{zitrin2009}, who note that this cluster 
has the largest known Einstein radius, $\theta_{e}\sim 55^{\prime\prime}$.  
\cite{zitrin2009} cite this large Einstein radius and the shallow surface 
mass distribution as further evidence that the cluster is disturbed, 
while \cite{limousin2012} characterize this as ``one of the most disturbed 
clusters presently known'' in their recent strong lensing analysis.  
We note that the lensing analysis of \cite{zitrin2009} assumed a redshift
$z \sim 2.5$ for the primary lensed system.  The recent redshift information reported in 
\cite{limousin2012} entails a shift in the normalization of the 
surface mass distribution $\kappa$, reducing the overall mass estimate for 
\macsc.  The analysis presented here only 
relies on the fact that lensing has located four mass peaks, and that
they show good agreement with the regions optically identified by \cite{ma2009}.

\cite{ma2009} provide the following interpretation of the four dominant
mass components in this merging cluster.
Subcluster A is the least massive and is likely falling back into the main
cluster from the NW, after having passed through once already.
Subclusters B and D are likely remnant cores that survived an initial 
encounter in a merger along an axis inclined much more toward the line of 
sight.  Subcluster C, which is the most massive component in \macsc, 
exhibits good X-ray/optical alignment and is most likely the disturbed core 
of the main cluster.  \cite{ma2009} also report the line-of-sight 
(optical) spectroscopically-determined velocities for subclusters A, B, C, and D
as $(v_{\mbox{\tiny A}}, v_{\mbox{\tiny B}}, v_{\mbox{\tiny C}}, 
v_{\mbox{\tiny D}}) = (+278^{+295}_{-339}, +3238^{+252}_{-242}, 
-733^{+486}_{-478}, +831^{+843}_{-800})$~km s$^{-1}$. 
We note the remarkably high line-of-sight velocity for subcluster B 
($\gtrsim 0.01~c$).

The Giant Metrewave Radio Telescope (GMRT) observed \macsc\ at 610~MHz
for a total of 4 hr \citep{vanweeren2009}.  
These GMRT observations reveal a powerful radio halo with a
spectral index $\alpha = -1.25$, as well as a 700~kpc wide substructure
identified as a radio relic by \cite{vanweeren2009}.  According to
\cite{vanweeren2009}, the progenitor of the radio relic is likely a
merger-driven shock wave within the cluster which has accelerated
electrons via the diffuse shock acceleration (DSA) mechanism.  The
radio substructure brackets the high-temperature regions of
the ICM (see Figure \ref{fig:xray_products}) and is oriented perpendicular 
to the merger axes and large-scale filament, which supports the relic scenario.

\cite{bonafede2009} performed high- and low-resolution observations 
of \macsc\ in full polarization mode with the Very Large Array (VLA),
at frequencies spanning 1.365--4.885~GHz.
They measure polarizations up to 20\% in the radio
substructure and find no sharp discontinuity of the $E$-vectors between
the large-scale halo and the putative relic, which is expected 
for a true relic \citep{clarke2006}.  \cite{bonafede2009} find that the 
observed lack of Faraday rotation does not agree with the expectation for a relic 
produced by a merger shock near the center of a cluster.  Lastly, there
is no steepening of the spectral index across the short axis of the
substructure, as would be expected of a radio relic following a
merger. Thus, \citeauthor{bonafede2009} argue that the substructure is
not a radio relic, but rather a bright, polarized filament connected
with the radio halo.  

The competing interpretations of the low frequency radio data both support the
hypothesis that merger activity has produced a relativistic, nonthermal component
to the ICM of \macsc, not seen in the \chandra\ X-ray observations implying it has
a relatively low density typical of radio relics/halos.

\section{Observations and Data Reduction}
\label{sec:obs}

\subsection{MUSTANG Observations}
\label{sec:mustang}

MUSTANG is a 64 element bolometer array operating at 90~GHz on the 100-m GBT
with a resolution of 8.$\!^{\prime\prime}$5.  For more information about
MUSTANG, refer to \cite{dicker2008}.\footnote{
\url{http://www.gb.nrao.edu/mustang/}.}

\begin{figure}[ht!]
    \includegraphics[width=3.25in]{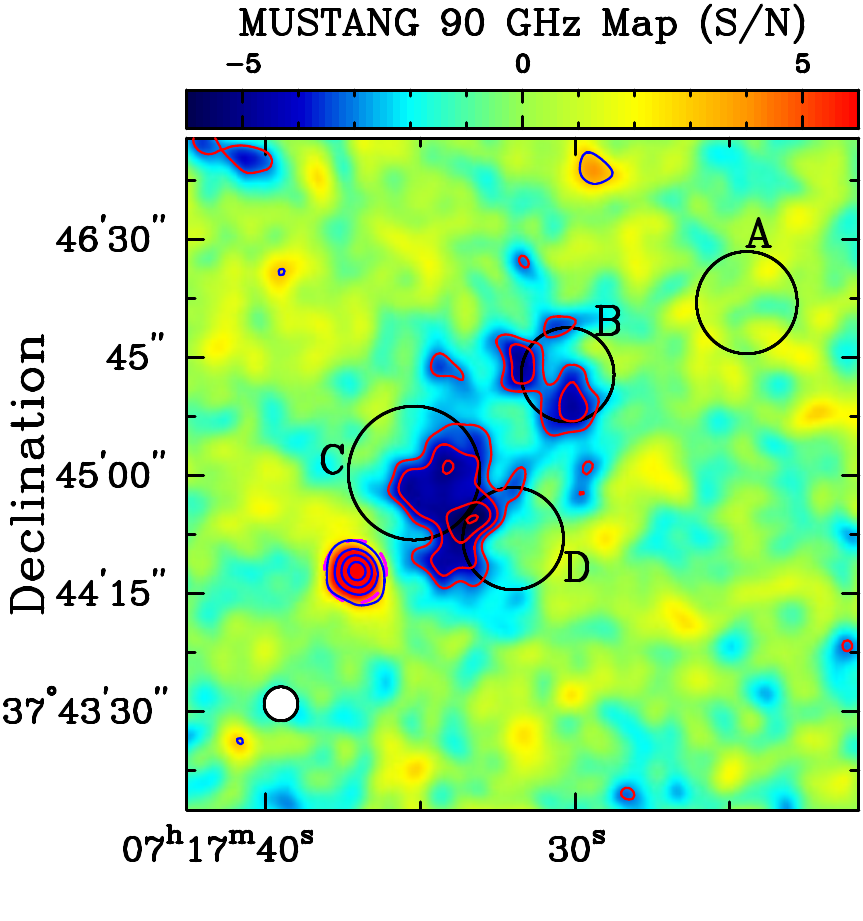}
    \includegraphics[width=3.25in]{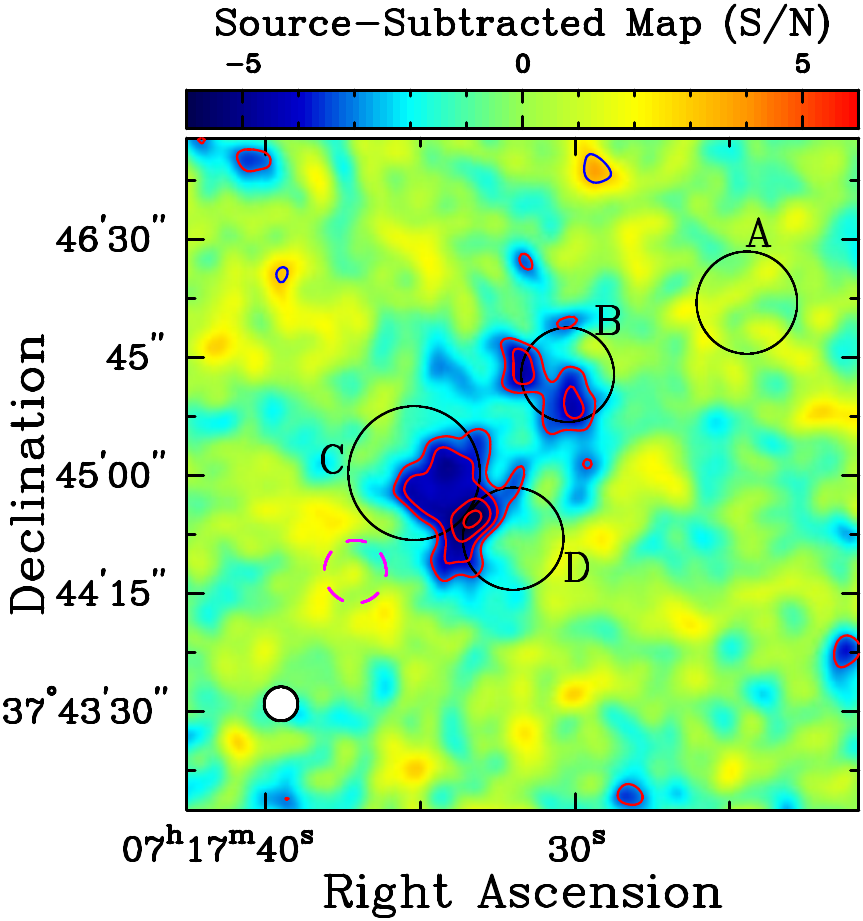}
  \caption{ 
    MUSTANG maps of \macsc\ with contours starting at
    $\pm3\sigma$, spaced at 1$\sigma$ intervals for the decrement (red) and at 
    3$\sigma$ intervals for the positive signal (blue).  The effective resolution
    of the map, which is 13$^{\prime\prime}$ due to the instrument beam 
    ($8.\!^{\prime\prime}5$ FWHM) and image smoothing with a 10$^{\prime\prime}$ 
    FWHM Gaussian, is depicted in the lower left corner of each map.
    The noise level in the inner $\sim 40^{\prime\prime}$ radial region is 34~\uJybm, 
    and is 53~\uJybm\ within the inner 3$^{\prime}$ radial region. 
    Due to our map-making process, features on angular scales $\gtrsim 1^{\prime}$ are 
    attenuated.  
    The largest extended SZE signal in this high-pass filtered view is located near subclusters 
    C and D, while a secondary SZE feature is also detected near B (see the text for details). 
    {\bf Upper:} MUSTANG map of \macsc\, before modelling and removal of the foreground source.
    {\bf Lower:} MUSTANG map with foreground radio source
    modelled and removed from the time-ordered data.
    Location of the removed source is marked with a magenta, dashed circle.
    Contours for the decrement are the same as above, but are now spaced at 1$\sigma$ 
    intervals for positive emission in the radio source-subtracted map (starting at +3$\sigma$).
    \label{fig:MUSTANGmaps}}
\end{figure}

Between 2010 October and 2012 February we observed \macsc\ for a total of 
15 hr on source over the course of 11 sessions.
Observations were carried out using an on-the-fly
scan strategy similar to that described in \cite{mason2010} and
\cite{korngut2011}.   For this, we move the telescope in a ``lissajous daisy'' 
pattern with a $3^{\prime}$ radius and slowly nutate the center of
the daisy pattern to increase the coverage. Seven different pointing
centers were used to further expand the coverage of the central part of the 
cluster. 
  
MUSTANG data were reduced and calibrated according to procedures described in
\cite{mason2010} and \cite{korngut2011}. 
The overall measurement errors and uncertainties in both the 
temperatures of Uranus and Mars, and the shape of the GBT's beam, limit our 
flux calibration to 10\% uncertainty in the absolute flux scale.
The data presented here are fully common-mode subtracted; that is, we remove
the instantaneous signal common to the detector array.  This has the 
benefit of removing atmospheric noise at the expense of attenuating signals from 
features with angular scales $\gtrsim 1^{\prime}$.\footnote{The transfer
function reaches half the power of that in the 10$^{\prime\prime}$--30$^{\prime\prime}$
range, where it is flat, at $\sim 60^{\prime\prime}$.  
See \cite{mason2010} for more details.}  
This limits the spatial dynamic range of the MUSTANG data presented 
here. 
Two largely independent implementations of the map-making pipeline
exist, and have different approaches for common-mode subtraction,
residual noise filtering, and data flagging. 
The features in Figure~\ref{fig:MUSTANGmaps} are recovered by both pipelines
with consistent brightness levels.

The MUSTANG map in the top panel of Figure~\ref{fig:MUSTANGmaps} has three main 
features, two due to the cluster and one due to a foreground elliptical
galaxy. The two extended SZE decrements have peaks of $-169 \pm 37$ and 
$-188 \pm 30$~\uJybm, and are detected at 4.6 and 6.2$\sigma$ 
significance at their peaks.  
The foreground galaxy, which is not co-spatial with the SZE in the
map, is resolved by MUSTANG and has an integrated flux density of 
$+2.8\pm0.2$~mJy (13.7$\sigma$ at the peak) and an extended 
shape of $14.4^{\prime\prime} \times 16.1^{\prime\prime}$.
The source is also detected at other wavelengths, including \chandra, GMRT, 
VLA (FIRST/NVSS), CARMA/SZA, and optical observations to name a few.

The radio sources detected in our maps at $>$3$\sigma$, 
such as the aforementioned foreground galaxy in the \macsc\ field, 
can be modelled and subtracted from our data.  
Provided the source has a significant detection, an initial 
estimate of the position, flux, and, if appropriate, shape of the source 
can be estimated from a first-pass map of the MUSTANG data.  
This model is then subtracted from the time ordered data before making a new,
point-source subtracted map (see lower panel of Figure~\ref{fig:MUSTANGmaps}).
If necessary, the source model convolved by the beam can be updated to better 
remove any residual flux after the first-pass point source removal.  
This method has the advantage over simply subtracting the source from 
the map in that the true shape and size of the source can be recovered 
while accounting for non-linearities of the map-making routine
(e.g., the influence of the source on the common mode subtraction and 
weight estimates).  One drawback to this method is that it subtly alters
the noise estimate in the S/N maps shown in Figure~\ref{fig:MUSTANGmaps}.
However, as we show in \S\ref{sec:mustang_analysis}, this has a negligible 
impact on flux estimates from the MUSTANG data.

\subsection{Bolocam Observations}
\label{sec:bolocam}

Bolocam is a 144 element bolometer array capable of operation at
either 140 or 268~GHz.  Operating from the 10-m CSO,
it provides resolutions of 58$^{\prime\prime}$ and 31$^{\prime\prime}$
at 140 and 268~GHz, respectively, over an 8$^{\prime}$ instantaneous field
of view.  A complete description of the Bolocam instrument can be found in
\cite{haig2004}.

We used Bolocam to observe \macsc\ for a total of 12.5 hours at
140~GHz (6 hr in 2010 February and 6.5 hr in 2010 October) and
for a total of 8 hr at 268~GHz (3 hr in 2011 September and 5
hr in 2011 November).  Due to the large instantaneous field of view,
these data have uniform coverage over the roughly $4^{\prime} \times 4^{\prime}$ 
region in which we are interested, with noise levels of 1.8 and
3.3~mJy beam$^{-1}$
for the 140 and 268~GHz data, respectively.  These data were reduced according 
to the procedures described in \cite{sayers2011} using the updated calibration 
model of \cite{sayers2012b}.  We briefly summarize the reduction below.

We used frequent observations of the nearby quasars to obtain pointing 
corrections accurate to 5$^{\prime\prime}$. 
We determined our flux calibration, with 5\% and 10\% 
uncertainties at 140 and 268~GHz, using observations of Uranus and
Neptune. We subtracted atmospheric noise from the data via a common-mode 
template and a time-stream high-pass filter with a characteristic
frequency of 250~mHz. The largest scales recovered after filtering (at either 
observational frequency) are 14$^\prime$. This noise subtraction also removes cluster
signal, and we determined the effective transfer function of our data
processing by reverse-mapping a simulated cluster profile into our
time-stream data and running it through the entire pipeline \citep[for
  details see][]{sayers2011}.  We note that the transfer functions
were determined independently for the 140 and 268~GHz data, and are
slightly different from each other.  

This data reduction results in a high-pass filtered image of the astronomical 
signal. Therefore, prior to comparing any model of the SZE signal to our data 
we must first convolve the model with both our point spread function (PSF) and the data-processing 
transfer function. Model fitting is otherwise straightforward, and the map 
noise is approximately white and is well described by a diagonal noise 
covariance matrix. Alternatively, we can also deconvolve the effects of the 
data-processing transfer function to obtain an unbiased image of the astronomical 
signal (aside from the effects of our PSF). 
Although the deconvolution amplifies the large scale noise in the image, it 
allows us to directly compute total flux densities from within an arbitrary 
aperture. In this paper, we exclusively use our processed data maps for all 
SZE model fitting to our Bolocam data, and we use our deconvolved 
data maps to obtain aperture-integrated fluxes which we compare to the model-fit
values. We note that in both cases, in order to 
fully account for any non-idealities in our noise, we determine all of our 
uncertainty estimates via 1000 statistically independent noise realizations.

\subsection{CARMA/SZA Observations}
\label{sec:sza}

The Sunyaev--Zel'dovich Array \citep[SZA; ][]{muchovej2007} is a subarray
comprising eight 3.5 m antennae in the Combined Array for
Millimeter-wave Astronomy (CARMA).  CARMA, including the SZA, is
located at an altitude of 2200~m in the Inyo Mountains of California.
The SZA is capable of observing in three modes: separately, paired
with the larger 6.1 and 10 m antennae of CARMA, and as part of the
full CARMA 23-element heterogeneous array \citep[see][for recent
observations with the full array at 90~GHz]{plagge2012}.  

Our observations use the 31~GHz SZA as an independent system in a
compact configuration sensitive to 1$^\prime$--6$^\prime$ scales with a
$10.\!^\prime5$ FWHM primary beam, which determines the field of view
\citep[for more details see][]{muchovej2007}.
The spatial filtering of the interferometer allows small scale positive 
point source emission to be separated from the
large-scale, negative SZE signal at these frequencies \citep[for an
example in a similar context see][]{reese2002}.  
The compact configuration provides two outrigger antennae (for a total of
13 of the 28 baselines) that probe $\sim 10\arcsec$ scales, though the 
rms noise level ($\sim 0.2$~mJy) is not low enough to measure the SZE, 
which is a factor of $\sim6\times$ lower at 31~GHz than at 90~GHz.
Our SZA 31~GHz observations are reduced using the pipeline described in 
\cite{muchovej2007}, and were calibrated using the \cite{rudy1987} model 
for Mars.  
The synthesized beam of the inner six elements of the SZA, which is
in a compact configuration to optimize cluster imaging on large scales,
was $118\times131^{\prime\prime}$ for these observations. We use the SZA
observations to constrain the bulk, arcminute-scale SZE decrement 
and to leverage the spectral indices of potential contamination by 
compact radio sources.  

\subsection{\chandra\ X-Ray Observations}
\label{sec:xray_obs}

The \chandra\ {\it X-ray Observatory} made two ACIS-I observations of \macsc\ 
for a total exposure time of 81~ks (ObsIDs 1655 and 4200).
We have reduced the data using CIAO version 4.3 and calibration
database (CALDB) version 4.4.5.  Starting with the level 1 events file,
standard corrections are applied along with light curve filtering and
other standard processing \citep[for details see][]{reese2010}.  A
wavelet-based detection algorithm is used to find point sources, which
is then used as the basis of our point source mask.
We discuss the data products we generate from our calibrated \chandra\ X-ray 
data in Section \ref{sec:xray}.

\section{Thermal Analysis of the ICM}
\label{sec:analysis}

We do not expect the assumption of spherical symmetry to be an accurate 
description of this merging cluster on subarcminute scales.  
This motivates the use of X-ray data to produce a two-dimensional template 
for the tSZE signal in our observations.
X-ray data can be used to infer electron pressure, $\Pe = \dene\kB\te$, 
the thermodynamic property to which the tSZE is linearly sensitive.
X-ray spectroscopy provides a measure of temperature \kB\te, while
X-ray imaging (i.e. surface brightness \sx) is proportional to the line of 
sight integral of the electron density squared, $\denesq$, times the 
X-ray cooling function $\Lamee$ ($\propto \te^{1/2}$).
Resolved temperature and pseudo-pressure maps, so-called because X-ray data
alone do not constrain the line-of-sight depth of the cluster, are common
in detailed X-ray studies of ICM thermodynamics 
\citep[see e.g.,][]{ma2009, russell2010}. 

The intensity of the tSZE is 
\begin{equation}
\Delta \Itsz = I_0 \, g(\nu,\te) \, y,
\label{eq:tsz}
\end{equation}
where the primary CMB intensity normalization is $I_0 = 2 (\kB
\Tcmb)^3 (h c)^{-2} = 2.7033 \times 10^8$ Jy sr$^{-1}$ ($=22.87~\rm Jy arcmin^{-2}$) and
\Tcmb\ is the temperature of the CMB.
Following \cite{carlstrom2002}, the function $g(\nu,\te)$ describes the frequency 
dependence of the tSZE.  We include the relativistic corrections of 
\cite{itoh1998} and \cite{itoh2004}.
The Compton-$y$ parameter in Equation~\ref{eq:tsz} is defined as
\begin{equation}
y \equiv \frac{\sigT}{\mec} \int \! \dene \kB \te \,d\ell 
= \frac{\sigT}{\mec} \int \! \Pe \,d\ell,
\label{eq:Comptony}
\end{equation}
where $\sigT$ is the Thomson cross section, \mec\ is the rest energy of the 
electron, the integration is along the line-of-sight $\ell$, and we have used
the ideal gas law ($\Pe = \dene\kB\te$).

\subsection{X-ray Data Products}
\label{sec:xray}

\begin{figure*}[ht!]
  \centerline{
    \includegraphics[width=3.25in]{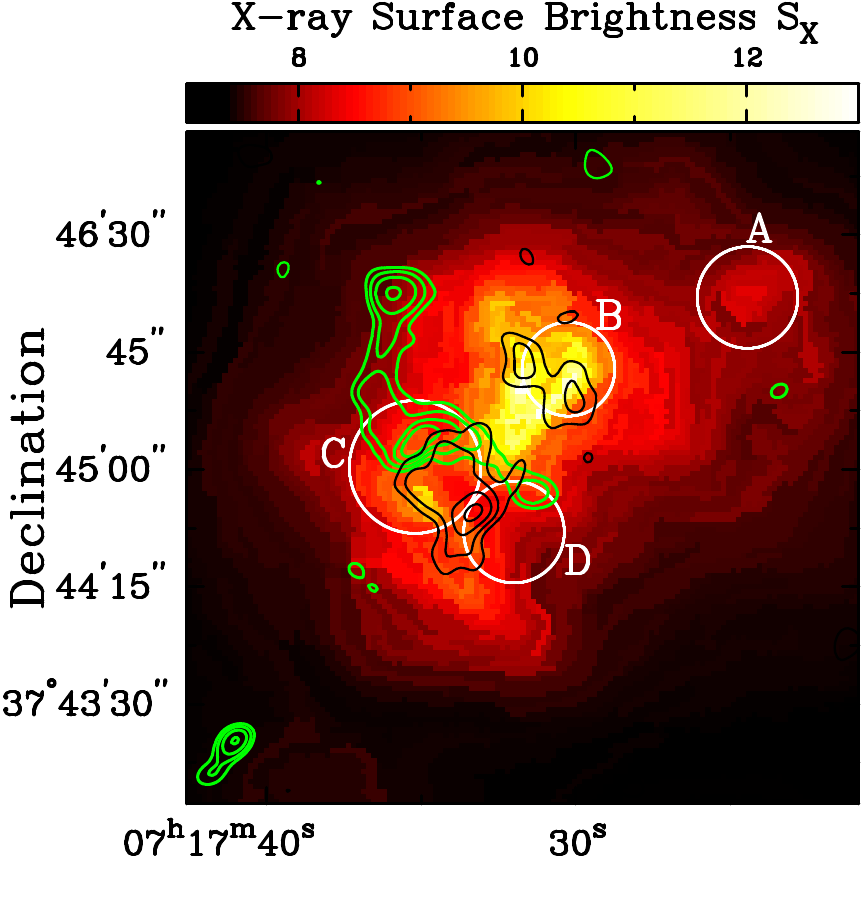}
    \includegraphics[width=3.25in]{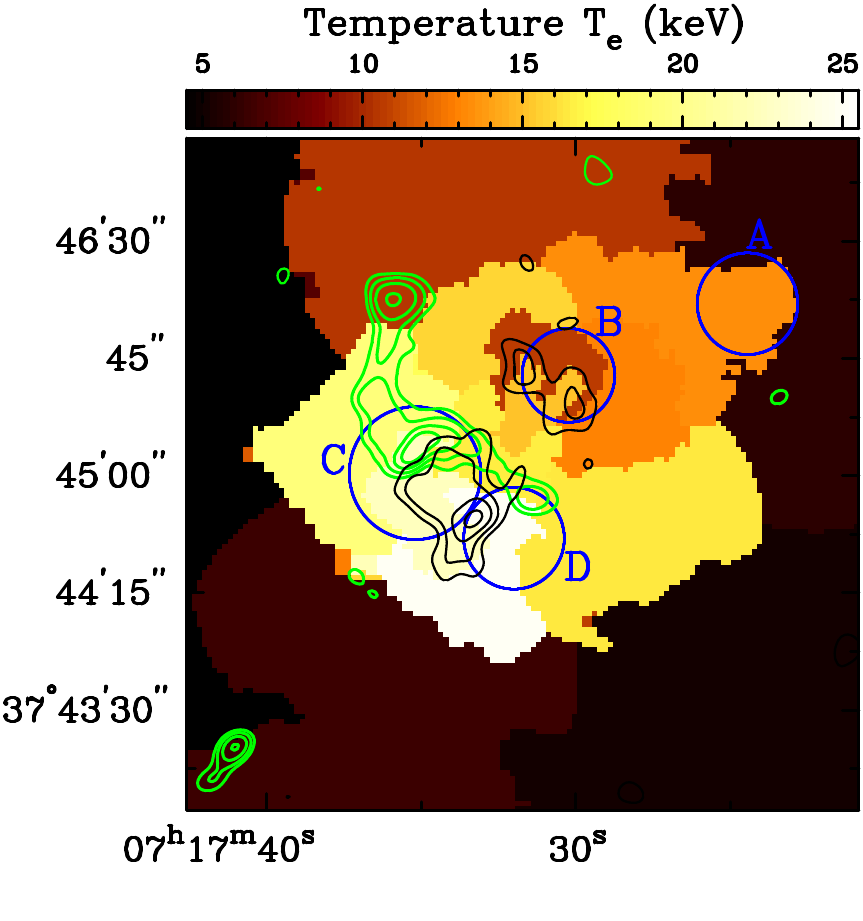}
  }
  \centerline{
    \includegraphics[width=3.25in]{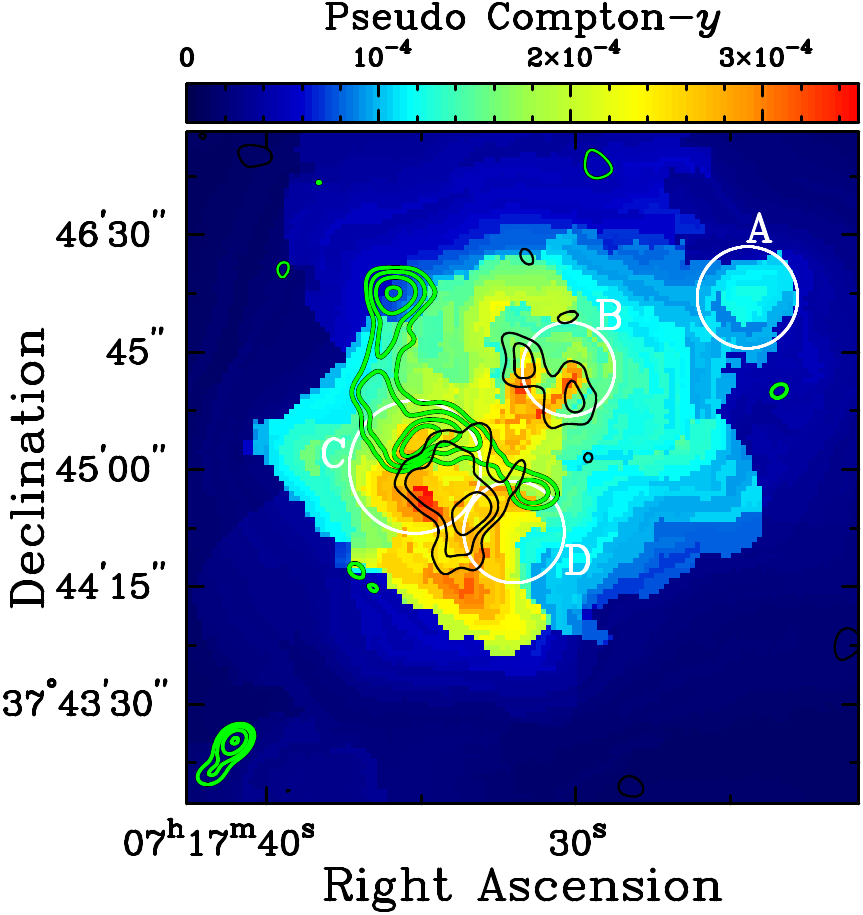}
    \includegraphics[width=3.25in]{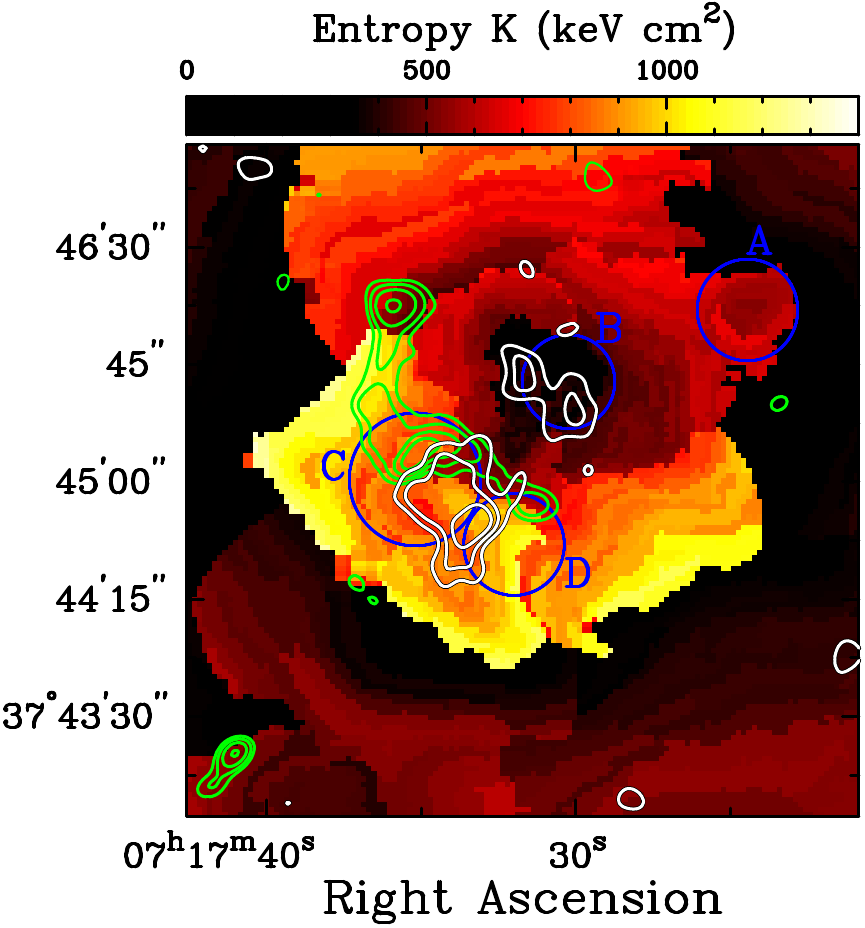}
  }
  \caption{Comparison between MUSTANG SZE detections and 
    the thermodynamic properties of \macsc\ inferred from \chandra\ X-ray data.
    MUSTANG SZE detections from lower panel of Figure~\ref{fig:MUSTANGmaps} are 
    overlaid (black contours, except for lower right panel).
    GMRT radio contours are overlaid in green (reproduced from \cite{vanweeren2009}).
    {\bf Upper left:} X-ray surface brightness map (shown on a scale 
    $\propto\sqrt{\sx}$). 
    {\bf Upper right:} X-ray temperature map.  The MUSTANG peak detections of the SZE 
    are located both where the hottest gas is found in the \chandra\ data (C and D) 
    and at the location of subcluster B.
    {\bf Lower left:} pseudo Compton-$y$ map.  This rescaling of the pseudo 
    pressure map aids in the interpretation of the SZE data, and serves as a 
    two-dimensional template for modelling the tSZE (Section \ref{sec:tsze}).
    {\bf Lower right:} map of the entropy parameter 
    $K = \kB \te / \dene^{2/3}$ in \macsc. 
    \label{fig:xray_products}}
\end{figure*}

To build the tSZE template, we first bin the reduced
\chandra\ data using {\it contbin} \citep{sanders2006}, which uses the
X-ray surface brightness to select regions of the image large enough to 
obtain a desired  S/N level.  For the temperature maps, we chose regions of the
X-ray surface brightness image with $\rm S/N > 45$.
Each bin provides independent measurements of surface brightness \sx,
temperature \kB\te, and metallicity $Z$.  For the spectral analysis to
determine \kB\te\ and $Z$, data extracted from each region are fit
jointly using spectra and response files produced from each individual
observation.  Regions containing point sources are excluded from the
extraction of the spectra and computation of the response files.
XSPEC is used to perform a joint spectral fit to both data sets 
over the 0.7--7.0~keV energy range,
linking temperature and metallicity between the data sets but allowing
the normalizations to vary.  The masked regions of the data product
maps are then filled in via a simple, bilinear interpolation scheme.

\begin{figure}[ht!]
    \includegraphics[width=3.25in]{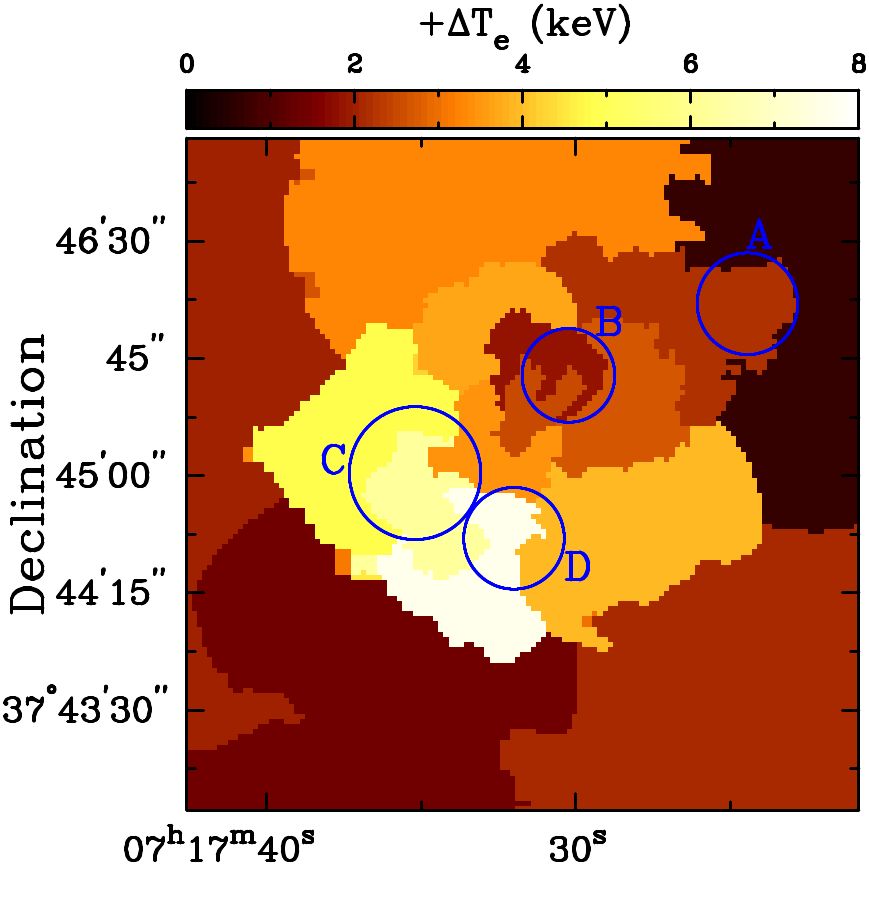}
    \includegraphics[width=3.25in]{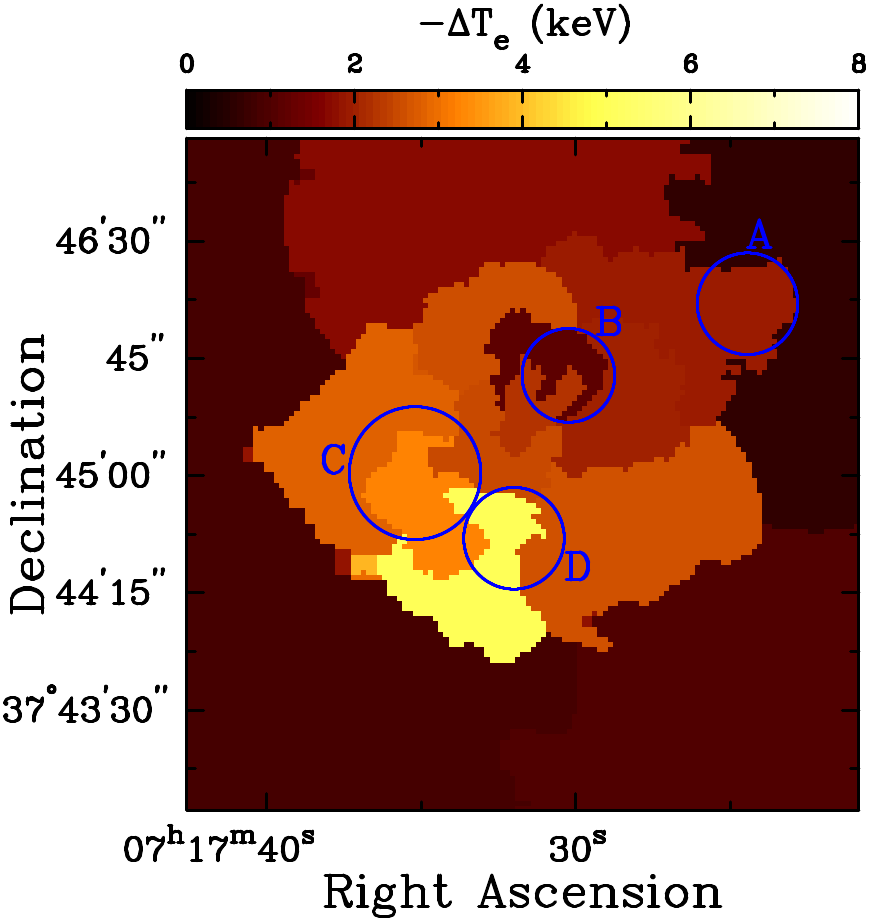}
  \caption{$\pm1\sigma$ statistical error bars for the
  temperature map presented in Figure~\ref{fig:xray_products}. Typical errors
  bars are $+3/-2$~keV for most regions, except for the hottest regions where
  subclusters C and D are interacting and the gas temperature is significantly 
  out of \chandra's energy band.
    \label{fig:Tx_errormaps}}
\end{figure}

Using the \chandra-derived \kB\te\ and $Z$ maps, we
compute the cooling function, $\Lamee(\te,Z)$, as a function of map bin,
and use it to compute a line-of-sight integrated pseudo-pressure map, where
\begin{equation}
\int \Pe \, d\ell \propto \kB \te \sqrt{\frac{4 \pi (1+z)^3 \sx \ell}{\Lamee(\te,Z)}}. 
\label{eq:pseudopressure}
\end{equation}
Here \sx\ (in $\rm counts~cm^{-2}~s^{-1}~sr^{-1}$) is the X-ray
surface brightness, and $\Lamee(\te,Z)$ (in $\rm counts~cm^{3}~s^{-1}$) 
contains the additional factor of $(1+z)^{-1}$
required by cosmological dimming, due to redshifting of the photon energy.
The factor $\ell$ is an effective electron depth of the
cluster along the line-of-sight, taken to be a single value over
the map.  Surprisingly, this is consistent with the average slopes found when 
using simple $\beta$-models to describe the density and pressure profiles 
\citep{sarazin1988,laroque2006,plagge2010}; a justification for this assumption may be found in the Appendix.  
We note that this does not imply pressure is constant along the line of 
sight, but rather that the average ratio of Compton-$y$ to $\sqrt{\sx}$ 
is approximately constant azimuthally.

We refer to this X-ray template for the tSZE, 
which is simply a rescaling of the X-ray pseudo-pressure map by $\sqrt{\ell}$, 
as a ``pseudo Compton-$y$ map.'' 
In constructing this, we have assumed Equation~\ref{eq:Comptony} can be approximated as 
$y = \sigT/(\mec) \int \Pe d\ell \approx \sigT/(\mec) \Pe \ell$. 
The key ingredient to building this template is the use of an SZE measurement 
on large angular scales to infer $\ell$.  
By using the integrated SZE flux, $\ysze = \int y \, d\Omega$, from the Bolocam 
140~GHz and SZA 31~GHz observations to
normalize the sum over the pixels in the pseudo Compton-$y$ map, 
we determine the median effective depth for the cluster to be 
$\ell \approx 1.4$~Mpc.  
The resulting maps are shown in Figure~\ref{fig:xray_products}.  
Adopting this median $\ell=1.4$~Mpc would impact the model
prediction for the tSZE in the MUSTANG observations on the 20\% level.
However, we effectively marginalize over the value of $\ell$ during the fits 
to the Bolocam data by allowing the normalization of the model to vary.

The comparison of X-ray pseudo-pressure to high-resolution observations of
the SZE is an important first step toward moving beyond simple, spherical 
models.  This is especially important for merging clusters, such as \macsc, 
that exhibit complicated thermodynamics and ICM distributions.
A similar approach has recently been applied by \cite{plagge2012} 
in the analysis of high-resolution SZE observations.
There are, however, several systematics that could affect the comparison
of X-ray pseudo-pressure to SZE data.
First, we have implicitly assumed that the temperatures in our {\it contbin} 
map are constant in each bin, both in the plane of the sky and along the line 
of sight, while errors on the temperatures can be as large as $\sim 25\%$
(see Figure \ref{fig:Tx_errormaps} and the discussion in Section \ref{sec:mustang_analysis}).  
Binning limits the spatial resolution in the resulting two-dimensional SZE template,
while the temperature distribution along the line-of-sight could affect
the normalization in each bin.
Second, density substructure effects such as clumping 
($\langle\denesq\rangle \geq \langle\dene\rangle^2$) 
and the presence of a cool core would bias our density 
estimates from the X-ray toward higher values, while the high luminosities 
of the cooler, denser clumps would bias the X-ray spectroscopic temperatures 
toward lower values. 

Using the same X-ray data products used for the pseudo Compton-$y$ map, we also
produce maps of the entropy distribution in this cluster.  
We adopt the entropy parameter 
$K = \kB\te \dene^{-2/3}~\rm [keV\,cm^2]$ 
commonly used in cluster astrophysics \citep[see e.g.,][]{cavagnolo2009}.
We approximate the (pseudo-)density 
\begin{equation}
\dene \approx \sqrt{\frac{4 \pi (1+z)^3 \sx}{\ell \Lamee(\te,Z)}},
\label{eq:pseudodensity}
\end{equation}
where $\ell$ is that inferred from the SZE data.
The maps in Figure~\ref{fig:xray_products} of the projected two-dimensional thermodynamic 
distribution imply that the highest pressure, hottest, and most entropic region 
is associated with the merger between C and D, while the remnant core of 
subcluster B exhibits a local entropy minimum.
Here the high local pressure substructure of B is due to its high density, 
suggesting this component, which has a high line-of-sight velocity, is relatively
intact and has probably approached or passed through the cluster with a high 
impact parameter.

\subsection{Thermal SZE Analysis}
\label{sec:tsze}

\begin{figure*}[ht!]
  \centerline{
    \includegraphics[width=2.33in]{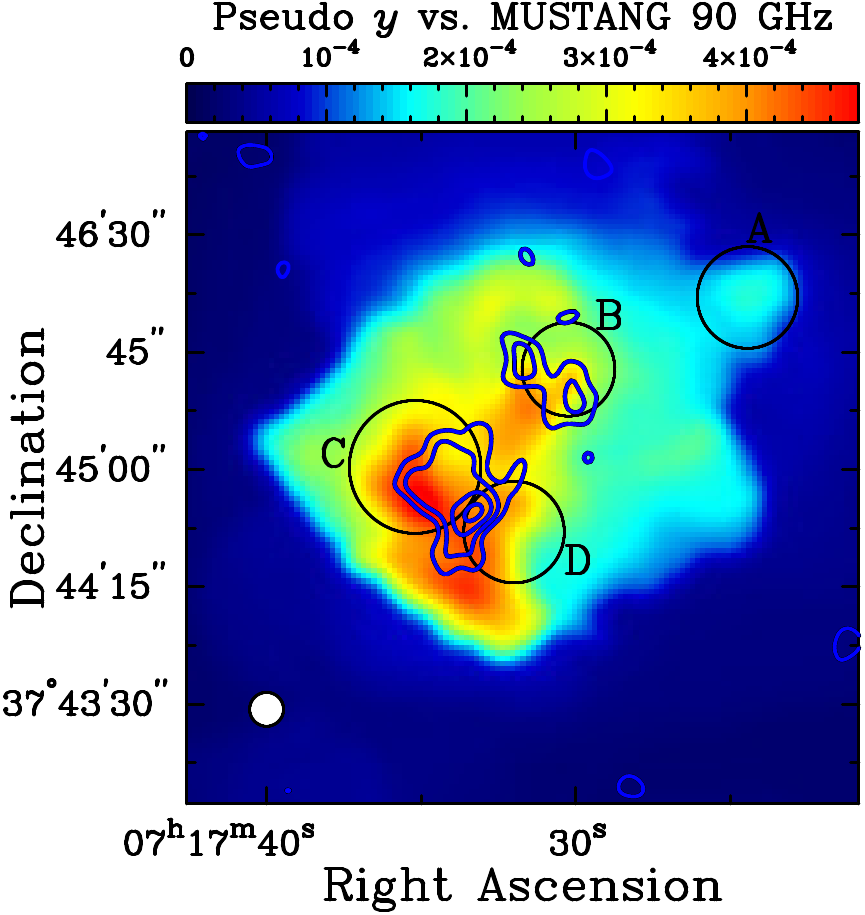}
    \includegraphics[width=2.33in]{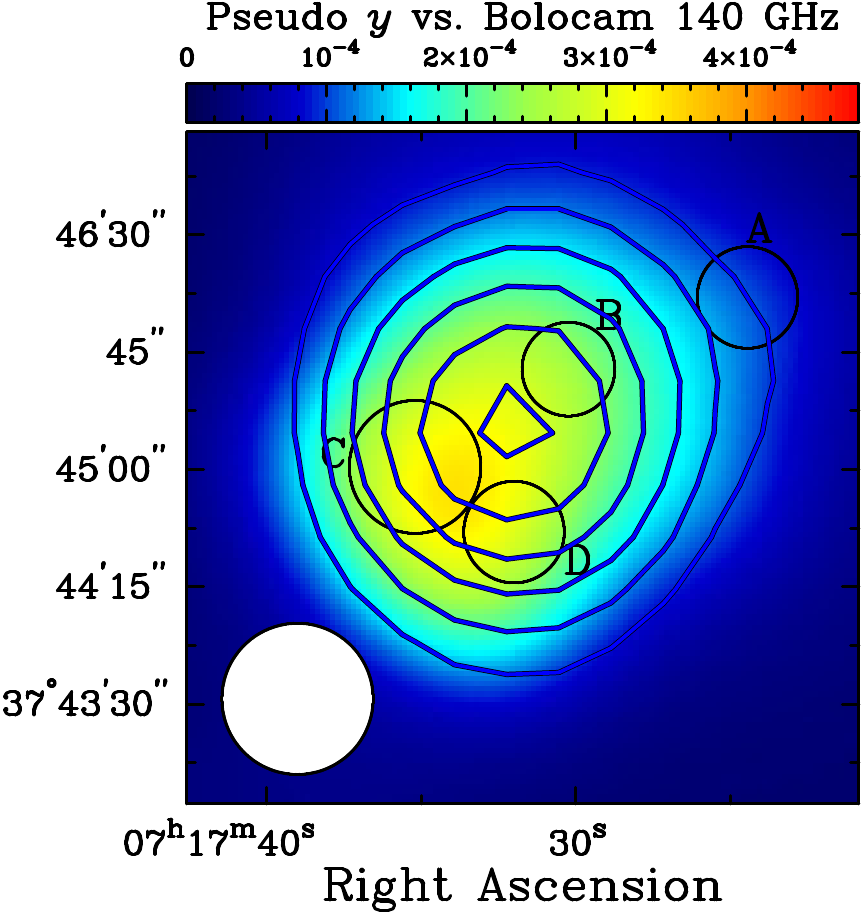}
    \includegraphics[width=2.33in]{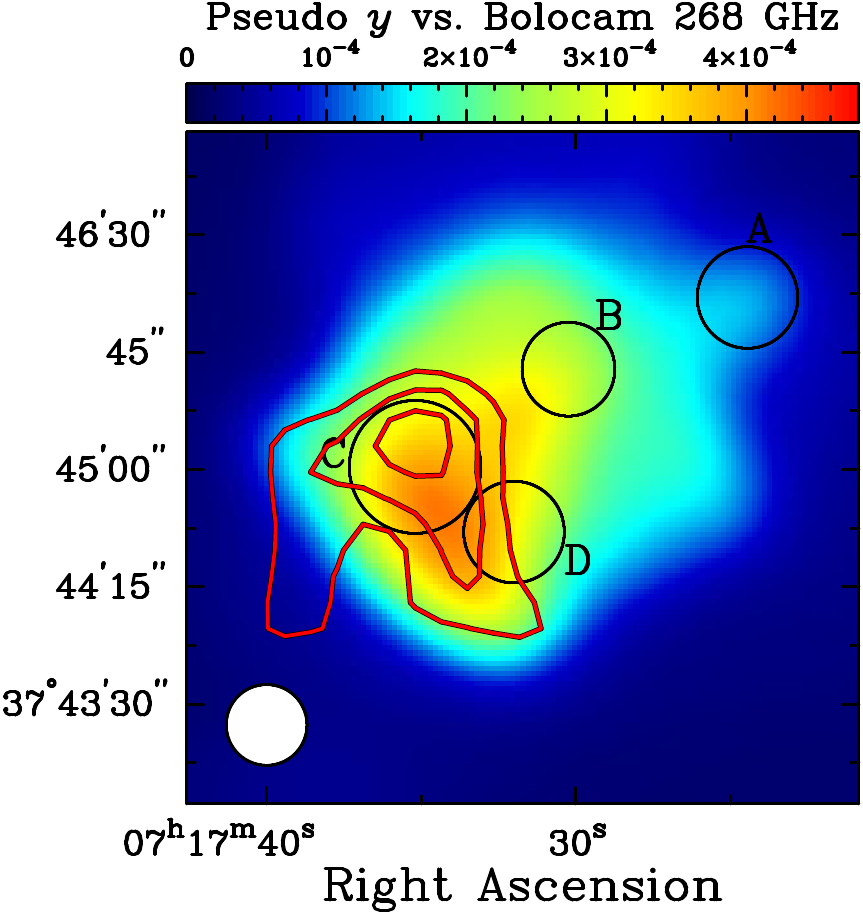}
  }
  \caption{Contours showing the SZE decrement (increment) for microwave
    observations below (above) $\sim$218~GHz.  Each observation is
    overlaid on the X-ray pseudo Compton-$y$ map smoothed to the resolution 
    of the instrument (FWHM depicted in the lower left corner of each panel).  
    The subcluster components identified by \cite{ma2009} are 
    labelled as in Figure~\ref{fig:lensing}.
    {\bf Left:} MUSTANG observation with point sources subtracted, overlaid 
    on a 10$^{\prime\prime}$ FWHM smoothed map.  
    Decrement contours are overlaid in blue at the 3$\sigma$, 4$\sigma$, 5$\sigma$, and 6$\sigma$ levels.  
    Note that the MUSTANG data overlaid are filtered with the instrument's
    transfer function, while the background image is simply smoothed to 10$^{\prime\prime}$.
    {\bf Middle:} decrement contours from the processed Bolocam 140~GHz data are 
    overlaid in blue at 5$\sigma$, 10$\sigma$, 15$\sigma$, 20$\sigma$, and 25$\sigma$ 
    on the 58$^{\prime\prime}$ FWHM smoothed pseudo Compton-$y$ map.  
    We note that the data are significantly shifted toward subcluster B, which would be 
    boosted by the kSZE by $\sim34\%$ assuming the optical velocity along the line 
    of sight, $v_{\mbox{\tiny B}} \approx 3200$~km~s$^{-1}$  
    \citep{ma2009}, is equal to the gas proper velocity.
    {\bf Right:} increment contours from the processed Bolocam 268~GHz observation, 
    overlaid in red at 3$\sigma$, 4$\sigma$, and 5$\sigma$ on the 31$^{\prime\prime}$ FWHM 
    smoothed pseudo Compton-$y$ map.  The SZE increment favors 
    the massive component C, which would be boosted by the kSZE up to 10\% 
    assuming the optically determined velocity along the line-of-sight, 
    $v_{\mbox{\tiny C}} \approx -700$~km~s$^{-1}$ \citep{ma2009}.  
    The total SZE from subcluster B would be suppressed up to $\sim68\%$ by the 
    kSZE of this high velocity component.  
   \label{fig:SZmap}}
\end{figure*}

Figure~\ref{fig:SZmap} shows the significance contours from each of our SZE 
observations, overlaid on the X-ray pseudo Compton-$y$ map smoothed 
to the resolution of each instrument.\footnote{Figure~\ref{fig:SZmap} is 
included only to facilitate the qualitative cross-comparison of our 
multi-wavelength SZE observations.  
These pseudo Compton-$y$ maps are merely smoothed to each 
instrument's resolution, and the transfer function is \emph{not} accounted 
for in the maps in Figure~\ref{fig:SZmap} in order to more accurately show 
the underlying pseudo Compton-$y$ prediction.
Figures~\ref{fig:mustangdiff} and \ref{fig:bolodiff} and the analysis 
in Sections \ref{sec:mustang_analysis} and \ref{sec:bolocam_analysis} do include the transfer function (i.e. the model 
there is processed in the same way as the data).}
Qualitatively, the maps at 90 and 140~GHz agree with the X-ray-derived
pseudo Compton-$y$ maps. 
There are two small-scale pressure peaks that are co-spatial with the MUSTANG 
detections of pressure substructure, and the 140~GHz data and 
the two-dimensional model (``the tSZE template'') broadly agree.
However, there are two main discrepancies:  the MUSTANG data show significant
levels of substructure -- particularly near B -- not predicted by the template, 
and the Bolocam 140~GHz data are shifted $\sim 20^{\prime\prime}$ toward 
subcluster B.  
This shift is significant and cannot be explained by pointing offsets;
including the $5^{\prime\prime}$ intrinsic uncertainty in the CSO pointing,
the Bolocam centroid is determined to $8^{\prime\prime}$ precision.

Looking at the 268~GHz Bolocam data (Figure~\ref{fig:SZmap}, right), we see a 
more profound disagreement than that seen in the observations of the SZE decrement.
The SZE increment from subcluster C clearly dominates the data, while no flux 
from B is apparent at $\rm S/N>1$.  
While the lack of agreement between the MUSTANG 90~GHz detection of subcluster B
and the non-detection at 268~GHz could be explained by filtering effects and the 
lower sensitivity in the 268~GHz Bolocam data, we note that the discrepancy 
between the Bolocam decrement and increment data \emph{cannot} be explained by the 
effects of signal filtering due to the atmospheric noise subtraction. 
The focal plane geometry, scan pattern, and atmospheric noise subtraction 
are identical in the Bolocam observations at both frequencies, 
and the effects of signal filtering are relatively mild and approximately 
the same at both frequencies.
While the noise level is much higher in the 268~GHz observation than that in the 
140~GHz observation, it is also clear from the SZE observations alone that flux 
is missing from component B at 268~GHz, while B is significantly brighter 
at 140~GHz than the X-ray data indicate it should be.

\begin{deluxetable*}{lcccccc}
\tablewidth{0pt}
\tablecolumns{7}
\tablecaption{Properties of SZE features observed by MUSTANG in \macsc.}
\tablehead{
Region 	 & \multicolumn{2}{c}{Peak Location (J2000)} 	& Peak $I_{90}$ 	& Peak $y$ 		& Integrated $Y$ 	& Temperature\tablenotemark{a} \\
	 &  R.A. 	& Dec. 				& ($\mu$Jy~bm$^{-1}$)  	&          		& $(10^{-12})$       	& (keV)
} 
\startdata 
\footnotesize
NW\tablenotemark{b}  	& 07:17:30.68 	& +37:45:38.1 	& $-169\pm37$ 	& $7.7\times 10^{-5}$ 	& $0.99\pm 0.2$ 	& 12.8$^{+2.1}_{-1.6}$ \\[.25pc]
NW src sub 		& 	  	&		& $-164\pm37$ 	& $7.5\times 10^{-5}$ 	& $0.98\pm 0.2$ 	& \\[.25pc]
NW src+relic sub	& 	 	&		& $-163\pm36$ 	& $7.5\times 10^{-5}$	& $0.98\pm 0.2$ 	& \\[.5pc]
SE\tablenotemark{c}  	& 07:17:33.95 	& +37:44:49.4 	& $-188\pm30$ 	& $9.0\times 10^{-5}$  	& $2.37\pm 0.3$ 	& 34.0$^{+11}_{-7.9}$\\ [.25pc]
SE src sub		& 	 	&		& $-207\pm33$ 	& $9.9\times 10^{-5}$	& $2.40\pm 0.2$ 	& \\[.25pc]
SE src+relic sub		& 	  	&		& $-206\pm34$ 	& $9.8\times 10^{-5}$	& $2.40\pm 0.2$ 	& 
\enddata
\label{tbl:szproperties}
\tablenotetext{a}{Temperatures reported here are from \chandra\ X-ray spectroscopic fits to the regions selected by 
MUSTANG at $>3\sigma$ over many beams.  These regions are smaller than the \cite{ma2009} regions, and
thus differ from the values reported in Table~\ref{tbl:ksz}. 
Due to the limited energy range available for X-ray spectroscopy ($\lesssim10$~keV),
the temperature of the hottest gas is poorly constrained.}
\tablenotetext{b}{Remnant subcluster core associated with \cite{ma2009} component B.}
\tablenotetext{c}{\chandra-detected hot-spot associated with \cite{ma2009} components C and D.}
\end{deluxetable*}

The properties of the SZE features observed by MUSTANG in \macsc\ are summarized in
Table~\ref{tbl:szproperties}.
In this table we provide the coordinates and integrated fluxes of
the SZE features in the MUSTANG observation.  For these,
we use the primary MUSTANG map (Figure~\ref{fig:MUSTANGmaps}, upper panel), 
the map with the foreground emission removed (``NW/SE src sub''; 
Figure~\ref{fig:MUSTANGmaps}, lower panel), and a map with all radio sources removed based 
on upper limits to their fluxes extrapolated from lower frequency radio observations 
(``NW/SE src+relic sub''; see the discussion below).  
For the two main SZE features, called the southeast (SE) and northwest (NW) features,
we provide estimates of the flux density from the raw map, from the map with the 
foreground radio source removed, and from the map with the foreground detected source 
and sources detected at lower frequencies and extrapolated conservatively to 90~GHz removed.
We note that our flux density estimates are consistent for all three maps, indicating that
the results are robust to radio source contamination.

Table~\ref{tbl:bulk_szproperties} reports the integrated Compton 
$\ysze=\int y \, d\Omega$ computed from model fits of the \cite{arnaud2010} pressure 
profile to the 31~GHz CARMA/SZA and 140~GHz Bolocam data.  
We find that, taken together, the flux in the NW and SE features, as sampled by 
MUSTANG's measurements (Table~\ref{tbl:szproperties}), 
account for $\sim 2\%$ of \ysze\ on large scales.

We describe below how we quantitatively compare the tSZE template to our 
SZE observations. 
For both MUSTANG and Bolocam the pseudo Compton-$y$ maps were first 
scaled to each instrument's observing frequency using the relativistic 
corrections of \cite{itoh1998} and assuming the temperatures shown  
in Figure~\ref{fig:xray_products}. 
Next, the maps were smoothed with the corresponding PSFs of each
instrument, and then filtered to account for the
signal attenuation in each instrument's data processing pipeline. 

\subsubsection{Modelling the tSZE in the MUSTANG Data}
\label{sec:mustang_analysis}

\begin{deluxetable}{lcccc}
\tablecolumns{5}
\tablewidth{0pt}
\tablecaption{Large Scale Thermal SZE properties of \macsc}
\tablehead{
Instrument & Frequency &  \multicolumn{2}{c}{Centroid (J2000)} &  $Y_{500}$\tablenotemark{a} \\
           &  (GHz)            &  R.A. & Dec.                          &  $(10^{-10})$
}
\startdata 
SZA     & ~31  & 07:17:30.4 & +37:45:25.9 &  $2.04^{+0.20}_{-0.17}$\\[.25pc]
Bolocam & 140  & 07:17:31.9 & +37:45:20.5 &  $2.44^{+0.50}_{-0.44}$ 
\enddata
\label{tbl:bulk_szproperties}
\tablenotetext{a}{Using $r_{500}$ from \cite{maughan2008}, where
  $r_{500}=1.36~\rm Mpc$ is the radius within which the average
  density is 500 times the critical density of the universe at that
  redshift.  This corresponds to $\theta_{500}=3.5^\prime$.  The
  values of $Y_{500}$ were determined by fitting the
  \cite{arnaud2010} pressure profile to the data, as in
  e.g., \cite{reese2012} and \cite{sayers2011}.}
\end{deluxetable}

\begin{figure*}[ht!]
  \centerline{
    \includegraphics[width=2.33in]{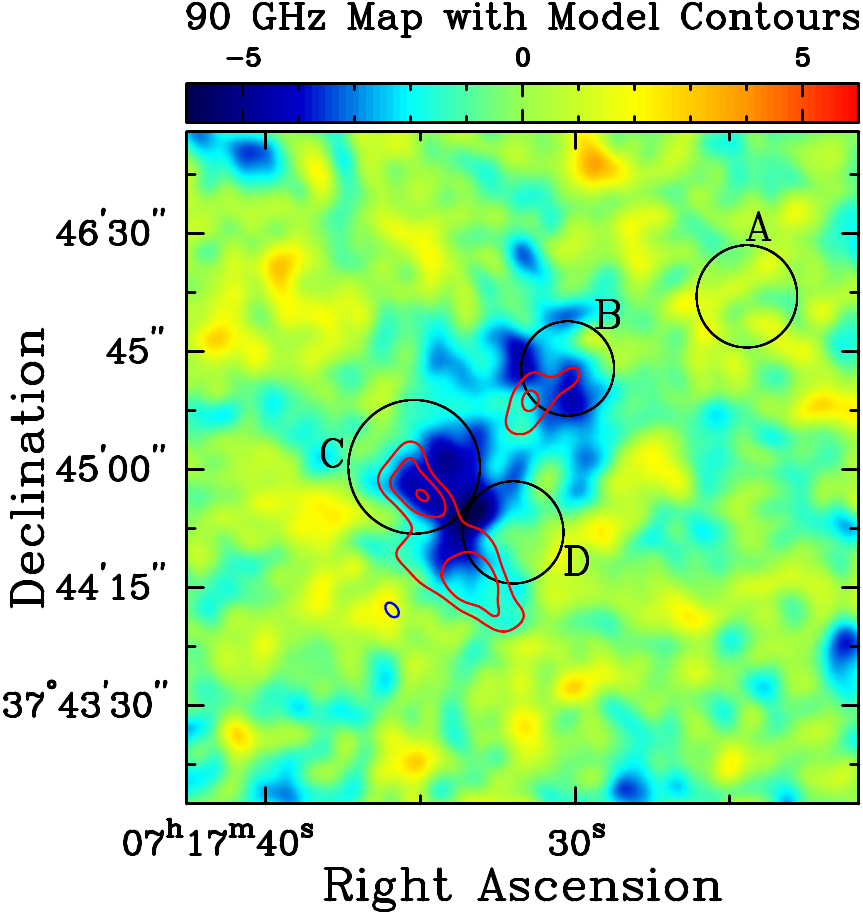}
    \includegraphics[width=2.33in]{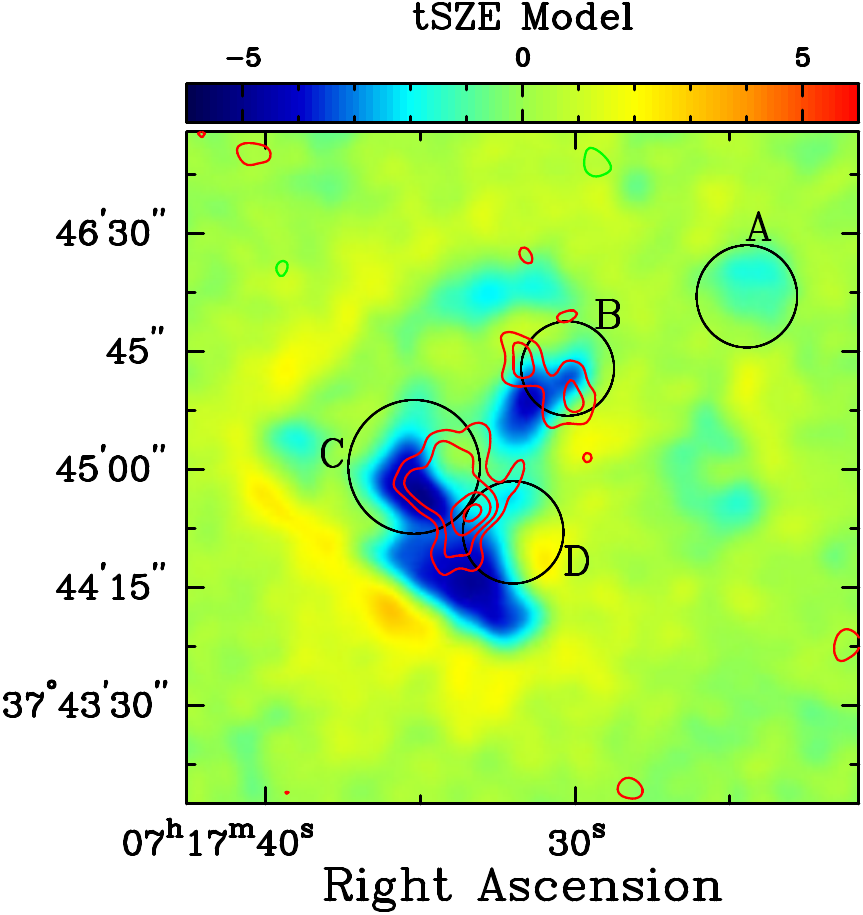}
    \includegraphics[width=2.33in]{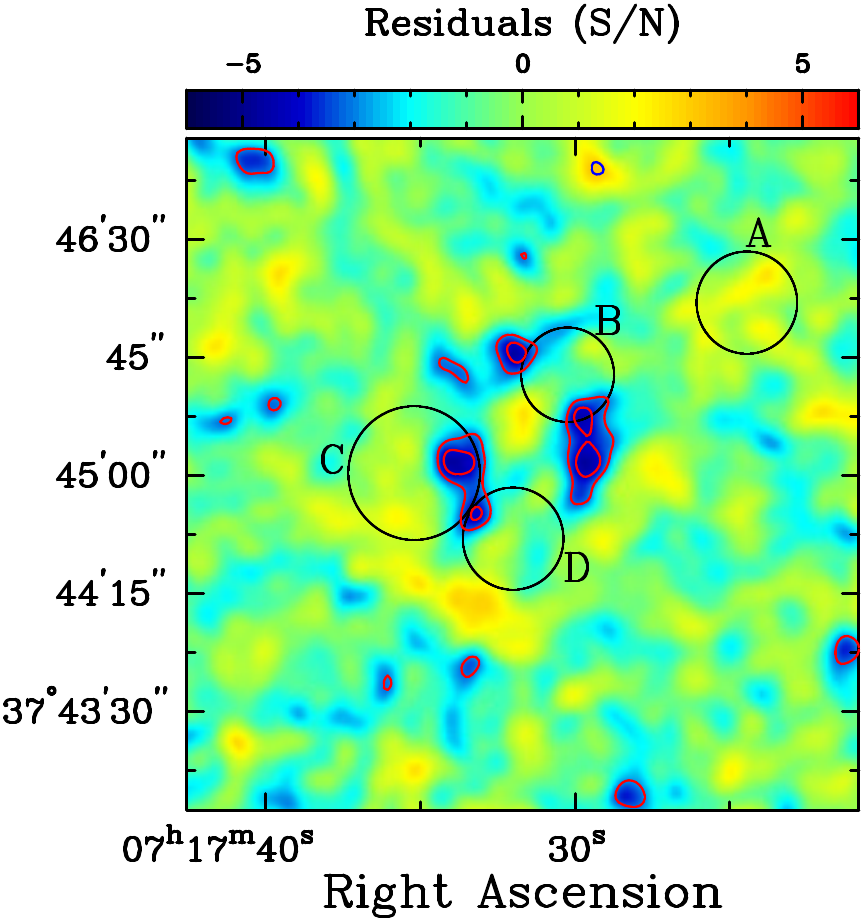}
  }
  \caption{ 
     {\bf Left:} MUSTANG map with foreground radio source subtracted (as in the lower
     panel of Figure~\ref{fig:MUSTANGmaps}).  The contours from the processed pseudo 
     Compton-$y$ model overlaid in red at the -102, -136, -170~\uJybm level.   
     {\bf Middle:} Pseudo Compton-$y$ tSZE template processed through the MUSTANG
     pipeline (corresponding to the contours from the left panel).  
     Contours from MUSTANG map with foreground radio source subtracted
     are overlaid at 3$\sigma$, 4$\sigma$, 5$\sigma$, and 6$\sigma$, 
     as in the lower panel of Figure~\ref{fig:MUSTANGmaps}.
     {\bf Right:} Residuals in the MUSTANG data after subtraction of the pseudo 
     Compton-$y$ map and radio point source model from the time ordered data (i.e.
     the scans used in map-making).  Contours are 3$\sigma$ and 4$\sigma$.
     The point source is cleanly modelled and subtracted, while the X-ray pseudo 
     Compton-$y$ map leaves significant ($>4\sigma$) residual flux on small scales, 
     potentially due to filtering effects in the model (e.g., discontinuities at the
     temperature region boundaries), the limited X-ray temperature 
     resolution in our \kB\te\ map, or the complicated line-of-sight structure of the 
     cluster.  See text for details.
    \label{fig:mustangdiff}}
\end{figure*}

We fit the radio source in the MUSTANG observation and subtract the cluster 
template as follows.  
First, we fit the emission from the foreground elliptical galaxy
(detected at $\rm S/N\approx13.7$) with an elliptical Gaussian. 
Accounting for our transfer function, the fitted elliptical
Gaussian is broadened to $14.4^{\prime\prime} \times 16.1^{\prime\prime}$ FWHM.
The cluster template is held fixed, and both the foreground elliptical galaxy and 
tSZE template were subtracted to produce the maps of the residuals shown in 
Figure~\ref{fig:mustangdiff}.

We also subtract the extended radio feature and compact
sources extrapolated from 610~MHz GMRT and 1.4 and 5~GHz VLA data,
using a constant power law extrapolation ($\alpha=-1.25$ for the extended
feature, and $\alpha=-0.7$ for the compact radio sources).
Included with this extrapolation was the detected foreground source.
Where associations can be made with FIRST \citep{white1997}, 
NVSS \citep{condon1998}, OVRO/BIMA \citep{coble2007}, or our CARMA/SZA data, 
we use the measured flux densities to help constrain the spectral index of the source.
We also tested subtraction of the extended radio feature using 
$\alpha=-1.4$.  In both cases, the extrapolated sources had a negligible impact.
Our spectral extrapolation should place a conservative upper 
limit on the radio source flux densities at 90~GHz.
Synchrotron sources, in general, show steeper spectra at higher frequencies 
due to radiative losses 
\citep[e.g.,][and references therein]{carilli1991, cotton2009}. 
For the inferred $\sim 3~\rm \mu G$ magnetic field of the extended halo/relic component 
\citep{bonafede2009}, the radiative lifetime of the electrons giving rise 
to emission at 90~GHz is $\sim 28$~Myr. 
If this population is older, then the spectrum is likely to be steeper 
than we have assumed.
Recently, \cite{marriage2011} measured a steepening on average 
of radio source spectra above 20~GHz.
We find that the contribution from the detected and undetected, extrapolated radio sources 
has a negligible impact on SZE flux measured at 90~GHz in these MUSTANG maps 
(see Table~\ref{tbl:szproperties}).  

After subtraction, we find a 4$\sigma$ residual flux decrement associated with the 
interactions of subclusters C and D, located between the strong lensing peaks.  
This southeast peak (SE) is likely due to the shock-heated gas produced in their 
merger, as indicated by the X-ray data and supported by the presence of extended 
radio emission to the north.
Restricting our X-ray spectral extraction to the region selected by MUSTANG at
3$\sigma$, containing $\approx 3000$ photon counts, we re-fit the X-ray data and 
find a temperature $\te=34^{+11}_{-7.4}$~keV (see Table~\ref{tbl:szproperties}), 
up from $\te=24.1^{+7.1}_{-3.5}$~keV for the contour binned map. 
If we further restrict our X-ray spectral extraction to the region selected by
the residuals in the right-hand panel of Figure \ref{fig:mustangdiff}, 
we find that we cannot place meaningful constraints on its temperature, 
despite having $\approx 1200$ X-ray photon counts.
This is presumably due to the temperature in that region being 
above \chandra's energy range.  
Since the tSZE is linear with temperature, such a high temperature would increase 
the expected signal over that in the pseudo Compton-$y$ map by $\sim 40\%$ from this 
region, and would account entirely for the tSZE residual we found.

We also find residual flux south of subcluster B.  This residual is of similar significance
as the residual near subclusters C and D.
Extracting $\approx 1000$ photon counts from the \chandra\ data for the region selected
by this residual, we find a temperature $\te=14.2^{+3.6}_{-3.2}$~keV.  This is
within the errorbars for the entire region selected by MUSTANG (see 
Table~\ref{tbl:szproperties}), while the region's X-ray surface brightness 
is much lower than that within region B.
We therefore interpret the residual south of region B to be largely an artifact of the 
subtraction of the pseudo Compton-$y$ template.  
While the X-ray information aids the interpretation of our high-resolution SZE data, 
the residual substructure could be due to a number of systematics, discussed below.

Foremost, the binning in the temperature maps produces large discontinuities
from region to region.  When high pass filtered, this can introduce a ringing
effect in the model image (see middle panel of Figure~\ref{fig:mustangdiff}).  
This seems to be the case for the region just to
the south of subcluster B.  A small, 3$\sigma$ feature appears to be boosted by
the positive ringing of an otherwise negative template.  
Improvements to the modelling and generation of pseudo Compton-$y$ templates 
will be explored in a future work including relaxed clusters with deeper X-ray data.
We note that the X-ray temperature constraints in our \kB\te\ map come
at the cost of reduced resolution in the resulting tSZE template.  Many
of the temperature bins are $\sim 1~\rm arcmin^2$ in area, while the resolution 
of the MUSTANG map is 13$^{\prime\prime}$.  Therefore, some of the larger 
temperature excursions due to shock heating on small scales could be missing
from our X-ray \kB\te\ map.

Errors in the X-ray derived temperatures as high as $\sim 25\%$
shown in Figure~\ref{fig:Tx_errormaps} give rise to two additional 
sources of uncertainty in the tSZE template.  The dominant one is simply that 
pressure scales linearly 
with temperature, while the additional error due to using the wrong temperature
when computing the relativistic correction to the tSZE at 90~GHz 
anticorrelates with temperature and is on the $\approx -3\%$ level for 
$\Delta \te/\te \sim +0.25$.
Additional sources of possible systematic error that could lead to discrepancies between
the X-ray prediction and SZE measurements are clumping and temperature
substructure (discussed in \S\ref{sec:xray}) and the breakdown of the 
assumption that the effective electron depth $\ell$ is constant 
(see Equation~\ref{eq:pseudopressure}).

\subsubsection{Modelling the tSZE in the Bolocam Data}
\label{sec:bolocam_analysis}

\begin{figure*}[ht!]
  \centerline{
    \includegraphics[width=2.33in]{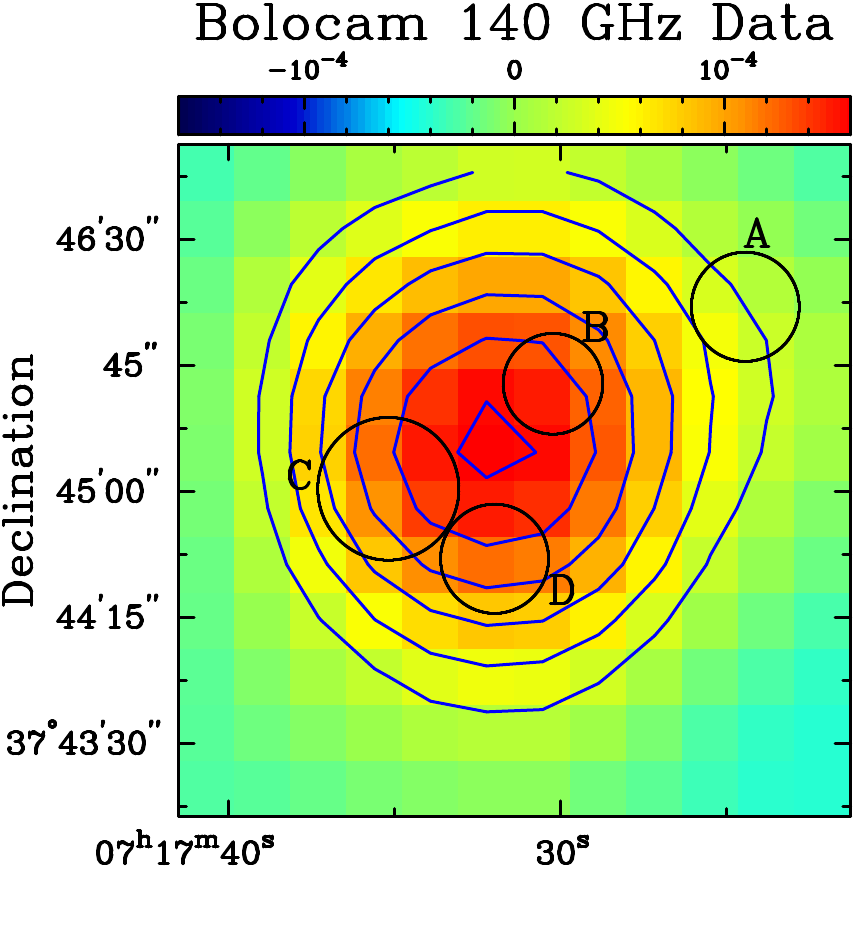}
    \includegraphics[width=2.33in]{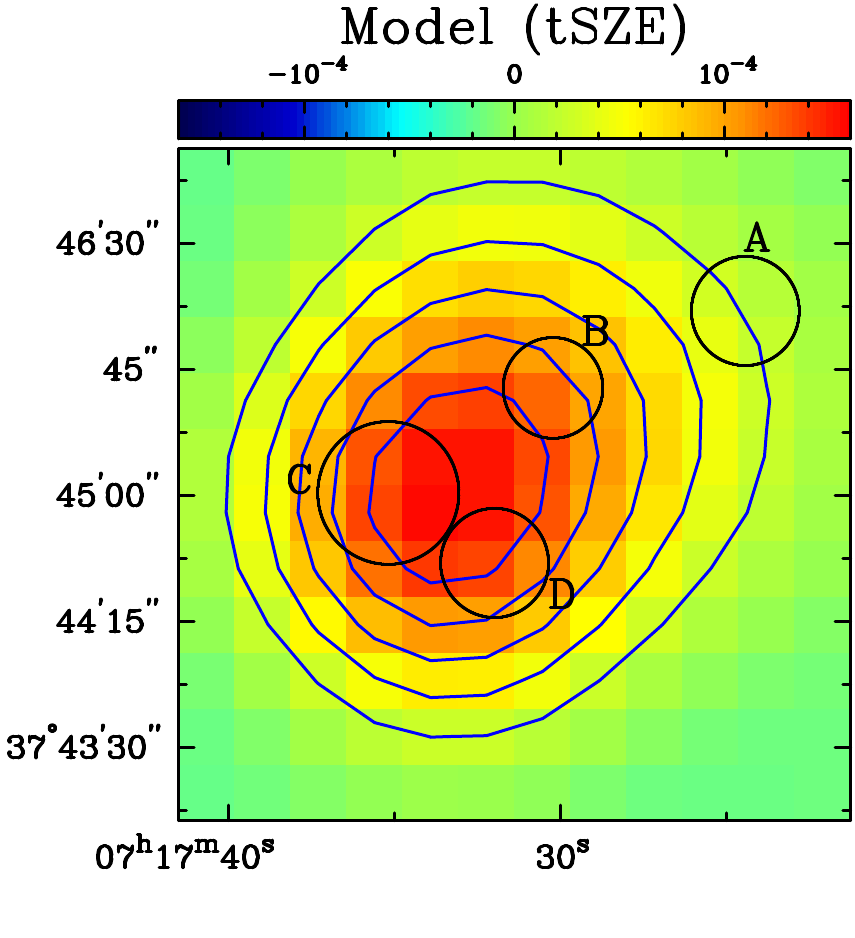}
    \includegraphics[width=2.33in]{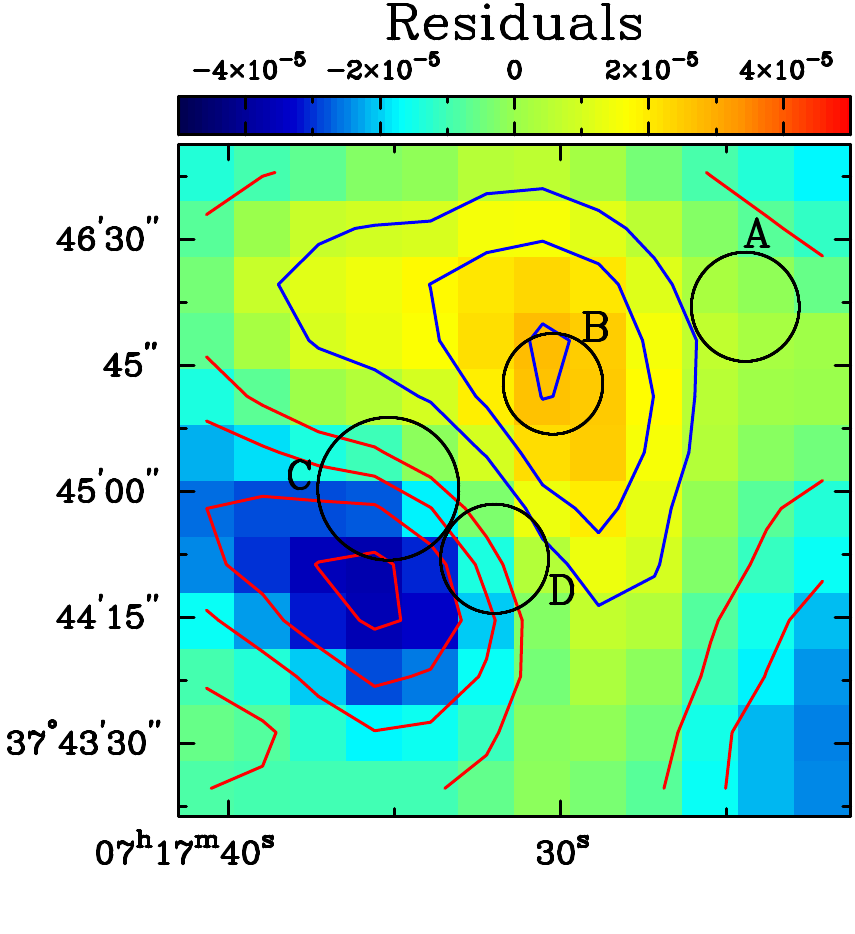}
  }
 \centerline{
    \includegraphics[width=2.33in]{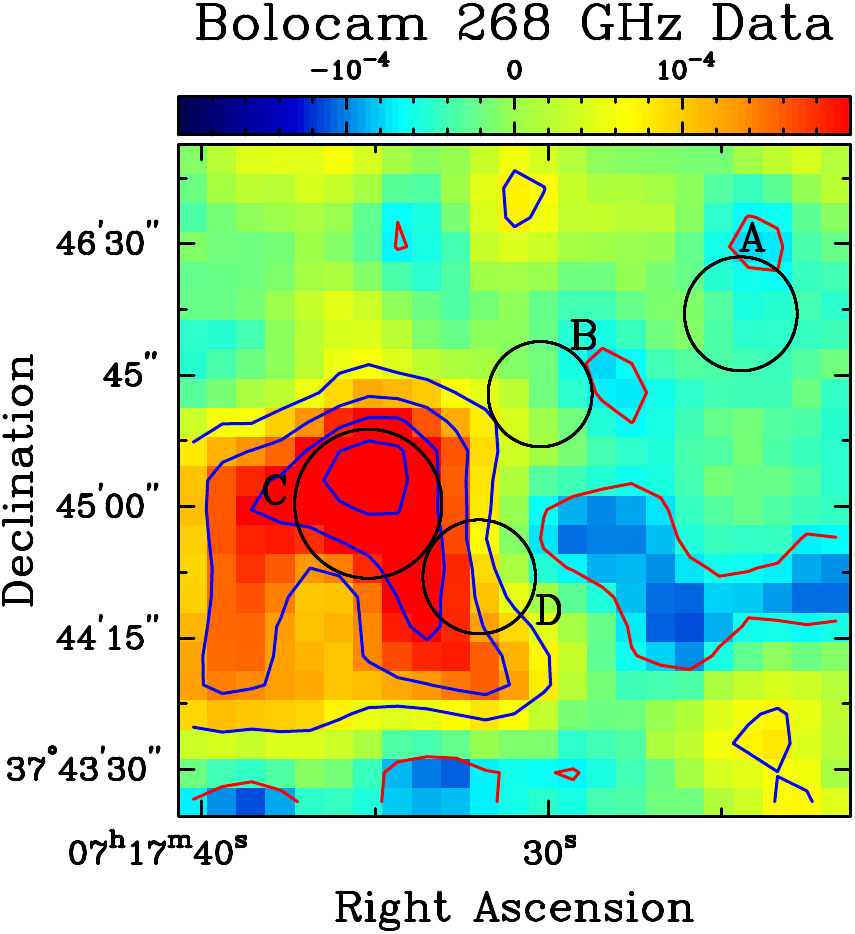}
    \includegraphics[width=2.33in]{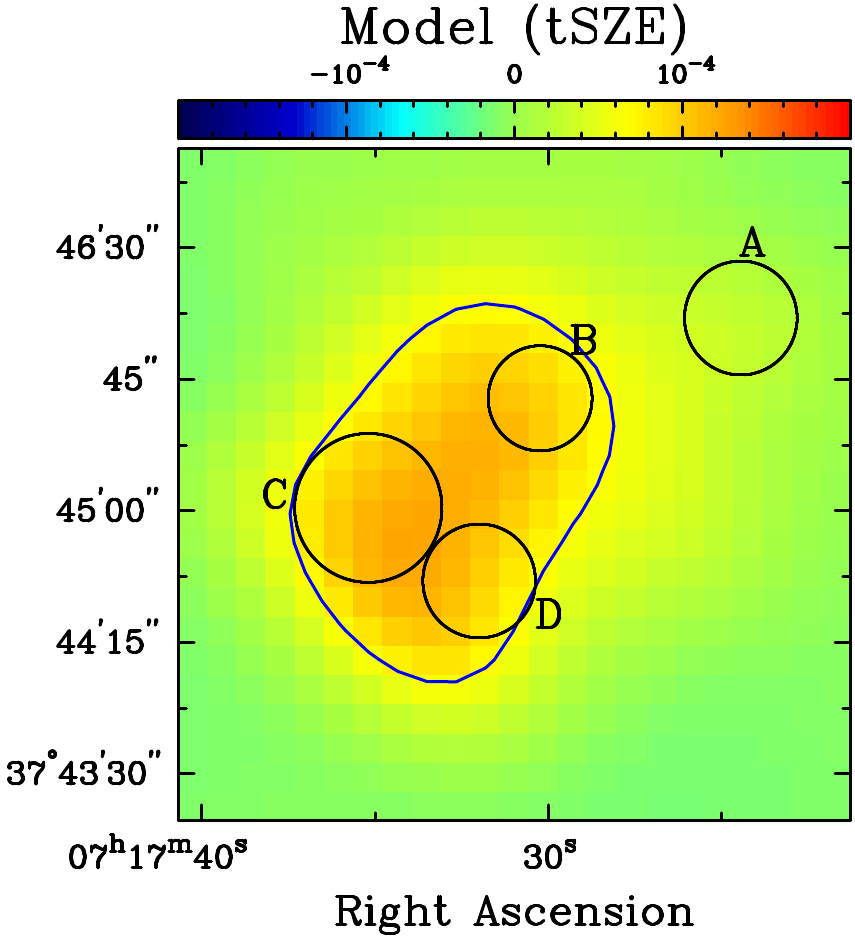}
    \includegraphics[width=2.33in]{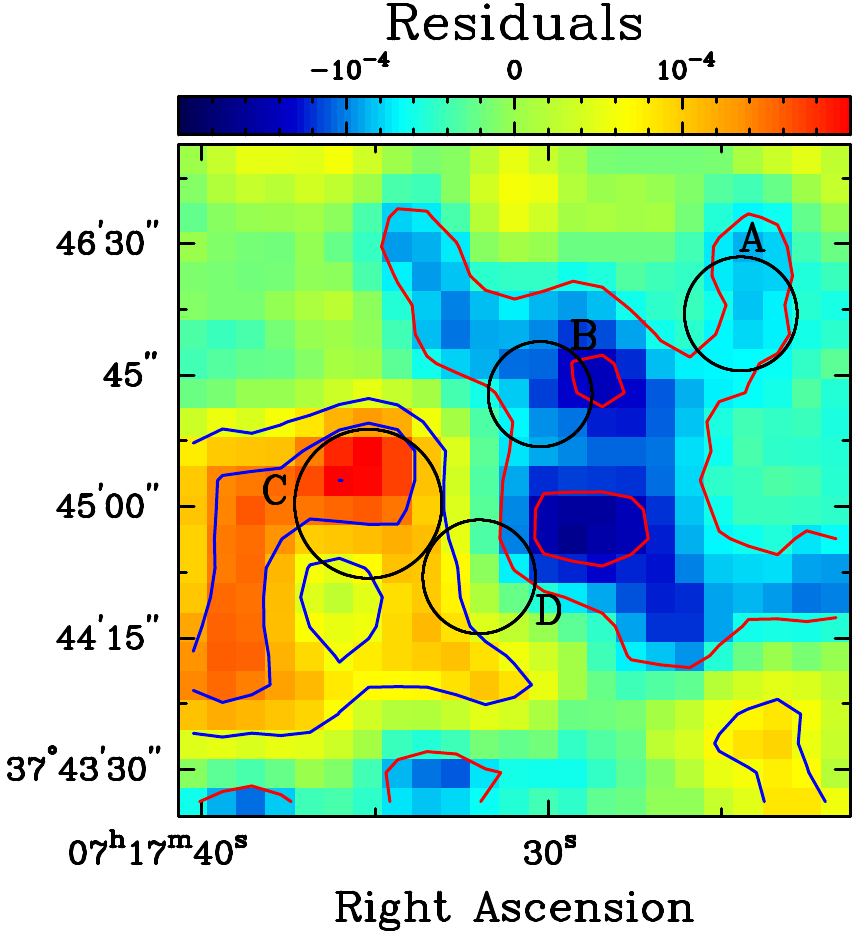}
  }
  \caption{
    Bolocam map, processed tSZE template (model), and residuals.  Upper panels are for the 140~GHz
    data, and lower panels are the corresponding plots for the 268~GHz data.
    {\bf Upper Left:} Bolocam 140~GHz map, smoothed with a 60$^{\prime\prime}$ 
    FWHM Gaussian. 
    Contours are overlaid at 3$\sigma$ intervals.
    {\bf Upper Middle:} Pseudo Compton-$y$ (tSZE) model at 140~GHz, processed through
    the same pipeline as the data, so that the filtering
    effects are the same for the model and data.  The normalization was 
    allowed to vary to find the best-fit value for the 140~GHz dataset. 
    Contours are overlaid at 3$\sigma$ intervals.
    {\bf Upper Right:} Bolocam 140~GHz residuals after the best-fit pseudo 
    Compton-$y$ model is removed. 
    Contours are overlaid at 1$\sigma$ intervals.
    Note the color scale was changed to enhance residuals.
    {\bf Lower Left:}  Bolocam 268~GHz map smoothed with a 30$^{\prime\prime}$
    FWHM Gaussian. Contours are overlaid at 1$\sigma$ intervals.
    {\bf Lower Middle:} Same as upper middle, but for the 268~GHz data. 
    Contours are overlaid at 1$\sigma$ intervals. 
    {\bf Lower Right:} Same as upper right, but for the 268~GHz data. 
    Contours are overlaid at 1$\sigma$ intervals.
    Note the color scale is unchanged for the 268~GHz plots due to the higher
    noise level.
    \label{fig:bolodiff}}
\end{figure*}

For Bolocam we fit the filtered pseudo Compton-$y$ map to our SZE data
using a least-squares method and allowing only the normalization of the map to 
float. In performing the least squares fit we weight the map pixels under the
assumption that the map noise covariance is diagonal, as in \cite{sayers2011}. 
However, since the diagonal noise covariance assumption is not strictly correct, 
we estimate our derived parameter uncertainties and fit quality via simulation 
using 1000 statistically-independent noise realizations, as described in more 
detail in \cite{sayers2011}. 
We note that by performing the fits in this manner,
our parameter uncertainties do not rely on our assumption that the noise
covariance is diagonal.

Due to limitations in the extent of the pseudo Compton-$y$
maps, which require accurate temperature determination from X-ray data, 
the fit was restricted to the inner $4^\prime \times 4^\prime$ region
of our Bolocam data.  We find best-fit normalizations of $1.043 \pm 0.092$
and $0.839 \pm 0.503$ at 140 and 268~GHz, indicating that the pseudo
Compton-$y$ maps are consistent with the total integrated SZE signal in the
Bolocam data at both observing frequencies.  This is expected for the 140~GHz
data since the pseudo Compton-$y$ map was normalized so it would have the same
\ysze\ as that found in fits to the Bolocam 140~GHz data. 
However, the fit quality is actually quite poor at 140~GHz, which we test
by computing the value of $\chi^2$ to determine the probability to exceed (PTE), 
which is the probability that we would obtain a fit with a larger value of 
$\chi^2$ due to a random noise fluctuation.\footnote{
Note that we quote values for the PTE based on the fraction of noise realizations
with larger $\chi^2$ values to identical fits, rather than quoting values based 
on the standard $\chi^2$ distribution. In \cite{sayers2011}, we found that the 
distribution of $\chi^2$ values from our noise realizations closely matched the 
theoretical distribution, and we again find that to be the case for the 140~GHz data. 
However, for the 268~GHz data presented here there is a noticeable difference, 
with our noise realizations producing a distribution of $\chi^2$ values slightly 
higher than the theoretical prediction. This discrepancy is likely due to the 
larger amount of atmospheric noise in the 268~GHz data.}
For the fit to the 140~GHz data, we find $\chi^2/\textrm{DOF} = 199/143$, 
with a $\textrm{PTE} = 0.1$\%, indicating that the pseudo Compton-$y$ map does 
not adequately describe our 140~GHz data (note that the fit quality is acceptable 
for the 268~GHz data, with $\chi^2/\textrm{DOF} = 624/575$ and $\textrm{PTE} = 22$\%).
Based on the opposite signs of the residuals in units of Compton-$y$ (see 
Figure~\ref{fig:bolodiff}), we identify a spectral dependence that suggests the 
presence of a kinetic contribution to the SZE signal, which we show in \S\ref{sec:ksze}
to improve the fit quality for both data sets.

\section{Inferred Peculiar Velocities}
\label{sec:ksze}

\begin{figure*}[ht!]
  \centerline{
    \includegraphics[width=2.33in]{macsj0717_Bolo140GHz_diff_map}
    \includegraphics[width=2.33in]{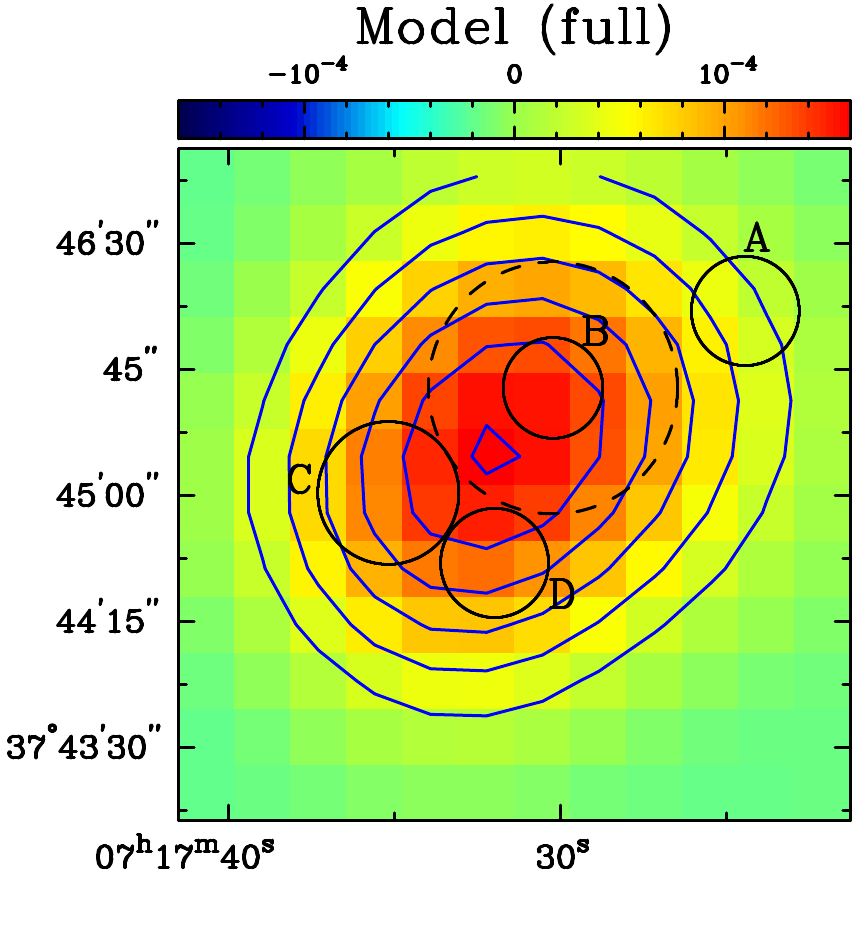}
    \includegraphics[width=2.33in]{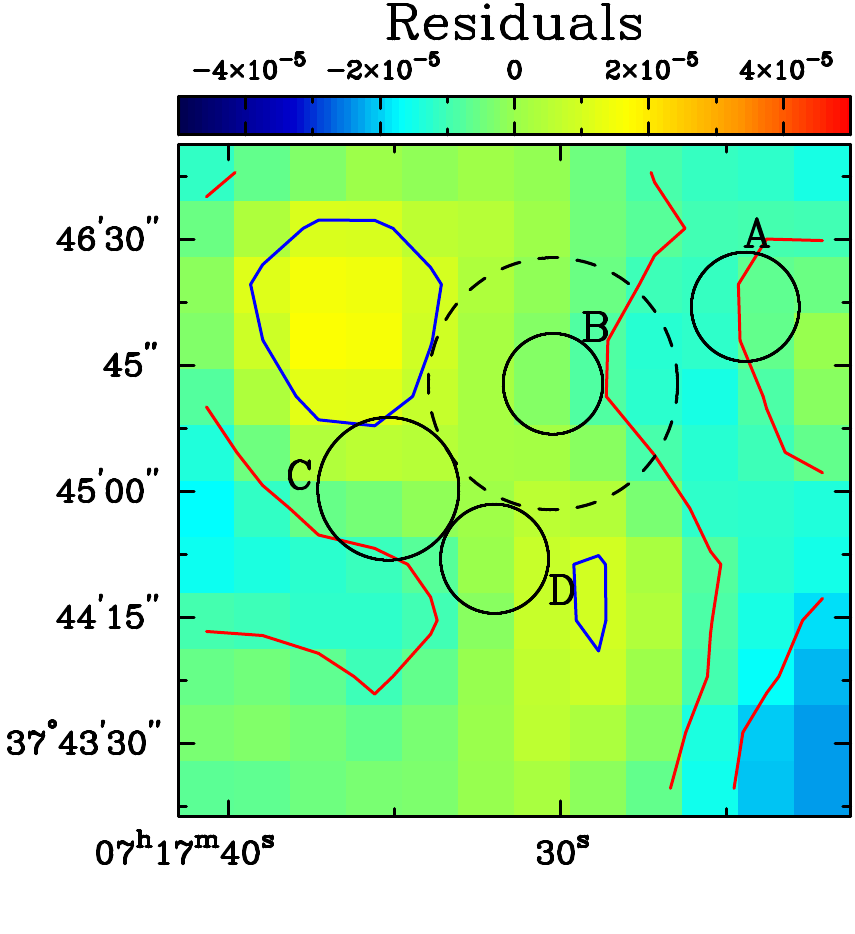}
  }
  \centerline{
    \includegraphics[width=2.33in]{macsj0717_Bolo270GHz_diff_map}
    \includegraphics[width=2.33in]{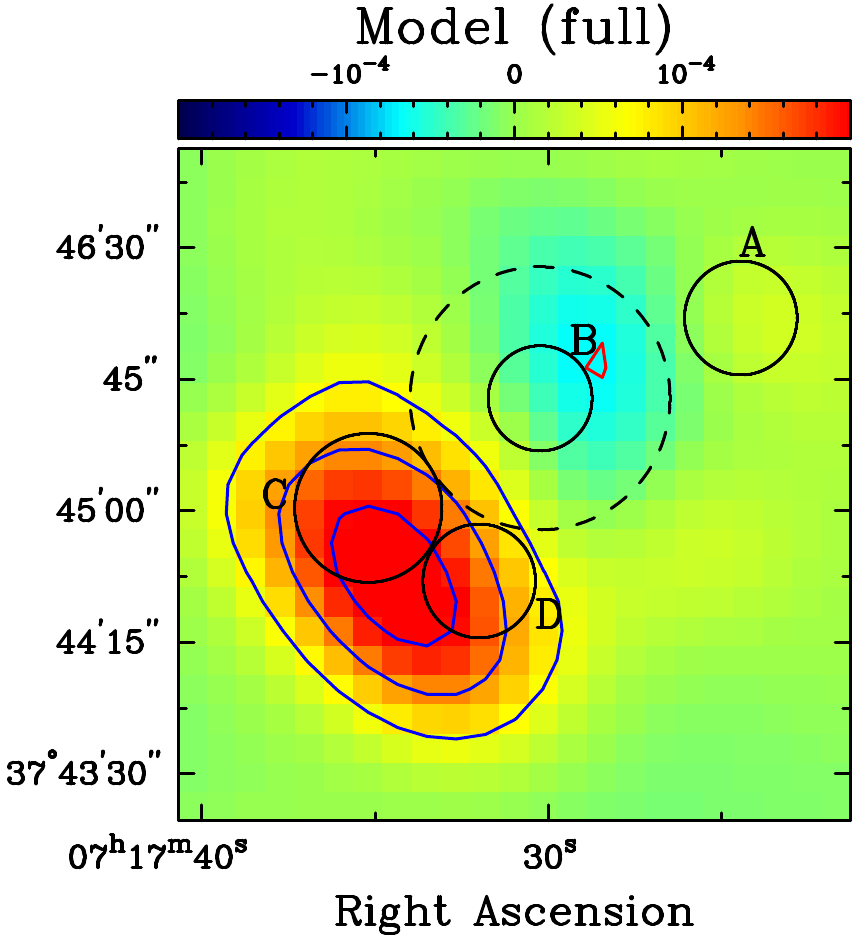}
    \includegraphics[width=2.33in]{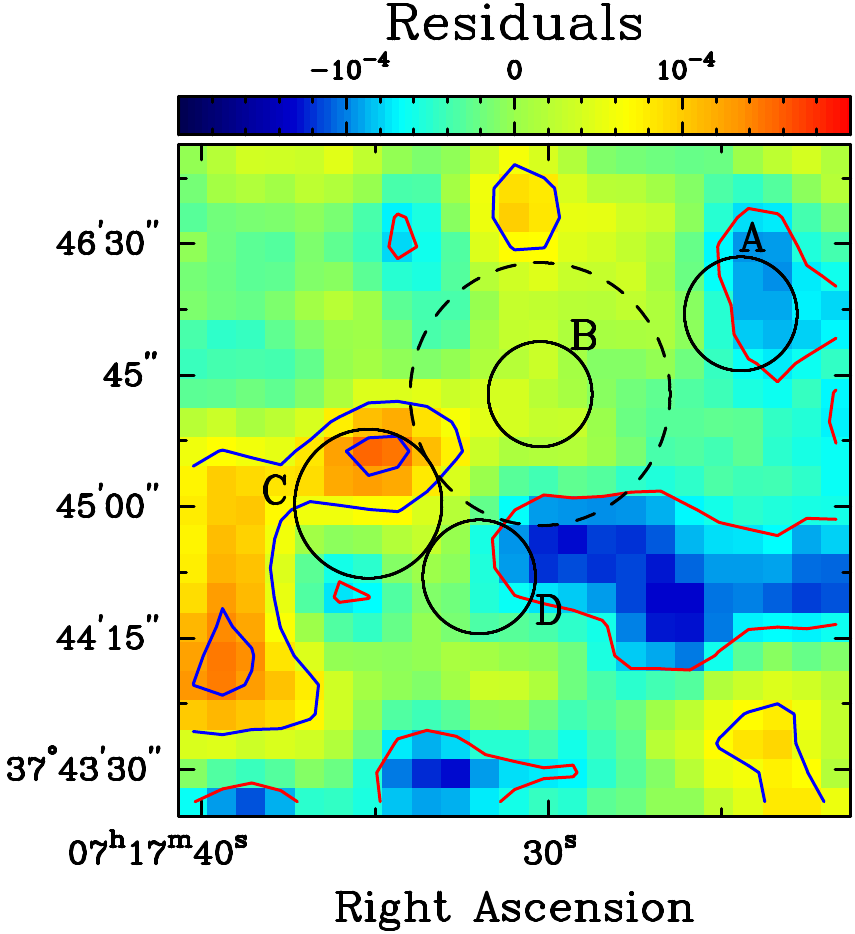}
  }
  \caption{
    Bolocam map, processed models (tSZE + kSZE Gaussian), and residuals.  
    Upper panels are for the 140~GHz data, and lower panels are the corresponding 
    plots for the 268~GHz data. Black, dashed circles
    in middle and right panels show location and size of added 90$^{\prime\prime}$ FWHM
    Gaussian model.
    {\bf Upper Left:} Bolocam 140~GHz map, smoothed with a 60$^{\prime\prime}$
    FWHM Gaussian. Contours are overlaid at 3$\sigma$ intervals.
    {\bf Upper Middle:} Model superposition of the best fit kSZE + pseudo Compton-$y$ template,
        processed through the same pipeline as the dataset it was used to fit.  
        The normalizations of both models were allowed to vary in order to find the best 
        fit for the dataset. 
        Contours are overlaid at 3$\sigma$ intervals.
    {\bf Upper Right:} Bolocam 140~GHz residuals with full model (kSZE + pseudo 
	Compton-$y$) removed. Contours are overlaid at 1$\sigma$ intervals, and the colormap 
	scale was changed to enhance residuals.
    {\bf Lower Left:} Bolocam 268~GHz map smoothed with a 30$^{\prime\prime}$
    FWHM Gaussian. Contours are overlaid at 1$\sigma$ intervals.
   {\bf Lower Middle:} Pseudo Compton-$y$ (tSZE) model at 268~GHz, processed through
	the same pipeline as the data, so that the filtering
  	effects are the same for the model and data.  The normalizations of both models 
	were freely varied.  Contours are overlaid at 1$\sigma$ intervals.
    {\bf Lower Right:} Bolocam 268~GHz residuals with full model (kSZE + pseudo 
	Compton-$y$) removed. Contours are overlaid at 1$\sigma$ intervals.
       Note the color scale is unchanged for the 268~GHz plots.
    \label{fig:bolodiff2}}
\end{figure*}

The residuals of the tSZE template fit, shown in Figure~\ref{fig:bolodiff},
indicate the reason for the poor fit quality of our tSZE template
to the Bolocam 140 and 268~GHz data. 
With the best-fit normalization given in \S\ref{sec:bolocam_analysis}, the pseudo 
Compton-$y$ map clearly under-predicts the 140~GHz signal near component B and 
over-predicts the signal 
near components C and D. Although the 268~GHz data are noisier, the opposite is 
true; for those data, the pseudo Compton-$y$ map over-predicts the 268~GHz signal 
near component B and under-predicts the signal near components C and
D. Motivated by these results, along with the known, large velocity of
the subcluster B, we consider the kSZE as a possible explanation 
of the opposite discrepancies between our 140 and 268~GHz data and the pseudo 
Compton-$y$ map. 
The intensity of the kSZE \citep{birkinshaw1999} is 
\begin{equation}
\Delta \Iksz = - I_0 \tau_e \, \frac{v_z}{c}  \frac{x^4 e^x}{(e^x - 1)^2}
[1 + \delta_{\mbox{\tiny kSZE}}(\te,x,v_z)],
\label{eq:ksz}
\end{equation}
where $v_z$ is the line-of-sight proper velocity (positive for a subcluster
receding from the observer), $\tau_e = \int \dene \sigT d\ell$ is the
electron depth, $x \equiv h \nu /(\kB \Tcmb)$ is the dimensionless
frequency, and $\delta_{\mbox{\tiny kSZE}}(\te,x,v_z)$ is a small
relativistic correction which we compute according to \cite{nozawa2006}.
It can be seen from Equation~\ref{eq:Comptony} that under the assumption
of constant temperature along the line-of-sight, $\tau_e \propto y / \te$.
For fixed Compton-$y$, the kSZE will be suppressed at higher temperature.

Since we lack precise knowledge of the shape of the possible kSZE signal
sourced by component B, we chose to describe it using a Gaussian profile
with a 90$^{\prime\prime}$ FWHM.\footnote{We varied the FWHM of the profile 
between 30 and 120$^{\prime\prime}$ and found a broad maximum in the fit quality at 
both 140 and 268~GHz, centered near 90$^{\prime\prime}$. We therefore left the 
FWHM  fixed at that value for the fit to the Bolocam data at both 140 and 268~GHz.} 
We then re-fit our 140~GHz data with a model consisting of the pseudo
Compton-$y$ map and the Gaussian source centered on component B, allowing
the normalization of each component to vary. The inclusion of this
Gaussian clump significantly improves the fit quality
($\chi^2/\textrm{DOF} = 153/142$ and $\textrm{PTE} = 24$\%), indicating
both that this Gaussian clump is justified statistically and
that the co-added model is adequate to describe our data. 
Jointly fit with the Gaussian clump, the normalization for the pseudo 
Compton-$y$ map fit to the 140~GHz data is $0.768\pm0.084$, indicating
the electron depth $\ell \sim 0.8$~Mpc (see \S\ref{sec:xray}).  
We then fit the same Gaussian model to our 268~GHz data, allowing the 
normalizations of it and of the pseudo Compton-$y$ map to float. 
For this fit, the normalization for the tSZE template fit to the 
268~GHz data is $1.78\pm0.68$.
We again find that including the Gaussian clump improves the fit quality
 ($\chi^2/\textrm{DOF} = 593/574$ and $\textrm{PTE} = 50$\%).
The combined models and residuals are shown in Figure~\ref{fig:bolodiff2}.

\begin{deluxetable*}{lccccc}
\tablecolumns{6}
\tablewidth{0pt}
\tablecaption{X-ray temperature and inferred peculiar velocities}
\tablehead{
Subcluster & \kB\te 
& $\Delta v$\tablenotemark{a}
& $S_{\mbox{\tiny140~\rm GHz}}$
& $S_{\mbox{\tiny268~\rm GHz}}$
& $v_{\mbox{\tiny \rm z}}$ 
\\
Name & (keV) 
& (km~s$^{-1}$)	
& (mJy)
& (mJy)	
& (km~s$^{-1}$)	
}
\startdata 
B (model)\tablenotemark{b}~
& 13.7$^{+2.1}_{-1.6}$  
& $+3238^{+252}_{-242}$ 
& $-19.5\pm2.0$
& ~$1.9\pm5.8$
& $+3600^{+3440}_{-2160}$ 
\\[.3pc]
B (nonpar)\tablenotemark{c} &  &
& $-17.0\pm2.1$
& $-0.5\pm7.5$
& $+4640^{+6720}_{-3360}$
\\[.3pc]
C (model)\tablenotemark{b}~
& 24.1$^{+7.1}_{-3.5}$  
& ~$-733^{+486}_{-478}$ 
& $-14.7\pm1.6$
& ~$18.5\pm7.5$
& $-3720^{+2960}_{-2480}$
\\[.3pc]
C (nonpar)\tablenotemark{c} &  &
& $-13.0\pm1.7$
& ~$17.7\pm8.5$
& $-4120^{+3760}_{-2880}$
\enddata
\label{tbl:ksz}
\tablenotetext{a}{Optical velocity from \cite{ma2009}.}
\tablenotetext{b}{Derived using fluxes computed from combined fit using a superposition of the
tSZE pseudo Compton-$y$ template and the Gaussian clump component described in the text.}
\tablenotetext{c}{Derived using non-parametric flux measurements from aperture photometry.}
\end{deluxetable*}

Using the results of our model fits to compute flux density, we attempt to fit for a 
tSZE+kSZE spectrum that can adequately describe subcluster B, the highest 
velocity member of \macsc.  
For comparison, we also fit the spectrum of subcluster C, since it is the most 
massive component, it has a lower known optical velocity ($\approx-700$~km s$^{-1}$), 
and it is detected at 5$\sigma$ in the 268~GHz data.
We also verify our model-dependent results using non-parametric, model-independent
estimates of the flux densities from the deconvolved images for each of these regions.
For all combinations of methods and regions, we use the integrated flux densities 
obtained from within 1$^\prime$ diameter apertures centered on subcluster B or C.
The values are listed in Table~\ref{tbl:ksz}.
We perform a grid search over the possible velocities, temperatures, and values
for \ysze\ in each region.  
We include the X-ray \kB\te\ constraints for each region when we compute
$\chi^2$ for each combined, relativistically-corrected thermal + kinetic 
SZE spectrum that can describe our measurements.  

Our joint constraints on velocity and \ysze\ for these regions, using both the 
model dependent and non-parametric fluxes, are shown in Figure~\ref{fig:ksz_conts}.
Also shown in Figure~\ref{fig:ksz_conts} are the 1 and 2$\sigma$ constraints on 
each parameter when marginalizing over the other (the $\Delta \chi^2 < 1$ and 
$\Delta \chi^2 < 4$ regions for projection to the axis of interest).
Note that the values of \ysze\ and peculiar velocity are anticorrelated, leading
to a wide possible range for both when fitting only two bands of SZE data.
Marginalizing over \ysze\ for each region, the median and 1$\sigma$ errors on velocity 
are reported in Table~\ref{tbl:ksz}.
The inferred kSZE+tSZE spectra are plotted in Figure~\ref{fig:ksz}.
We find that the probability of $v_z\leq0$ is 2.1\% for the 
fits to the model-derived flux density for subcluster B, and 3.4\% for the non-parametric
flux density estimates.  For subcluster C, we find the probability of $v_z\geq0$
is 15.7\% for the model flux density estimates, and 19.3\% for the nonparametric flux 
density estimates.
All of our peculiar velocity estimates agree with the  
\cite{ma2009} optical estimates to within 1$\sigma$.

The possibility of using the CARMA/SZA 31~GHz data, which should be relatively
insensitive to the kSZE contribution, to better constrain the tSZE component 
was considered.  However, the synthesized beam 
in our CARMA/SZA 31~GHz observations was over 2$^\prime$, while we are fitting 
components at the arcminute scale, and would thus make the results difficult 
to compare directly.

\begin{figure*}[ht!]
  \centerline{
    \includegraphics[width=3.5in]{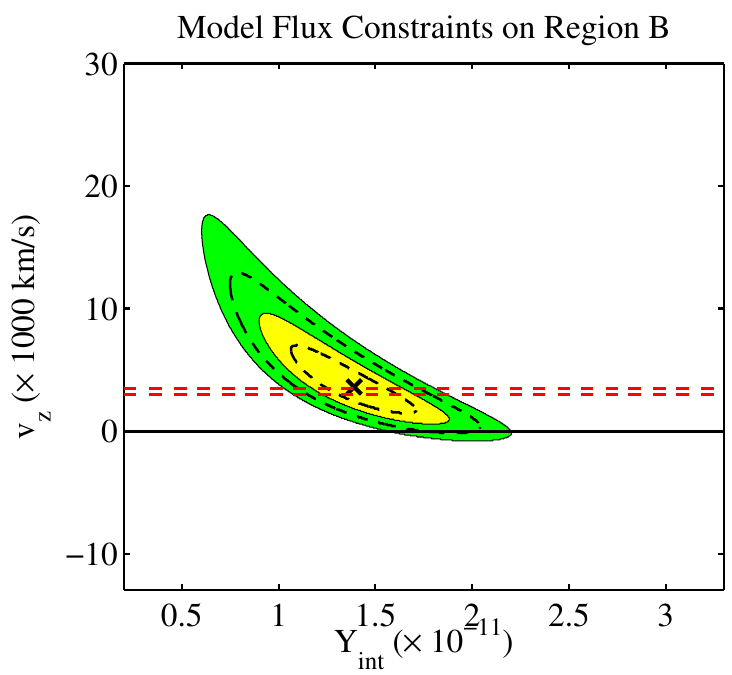}
    \includegraphics[width=3.5in]{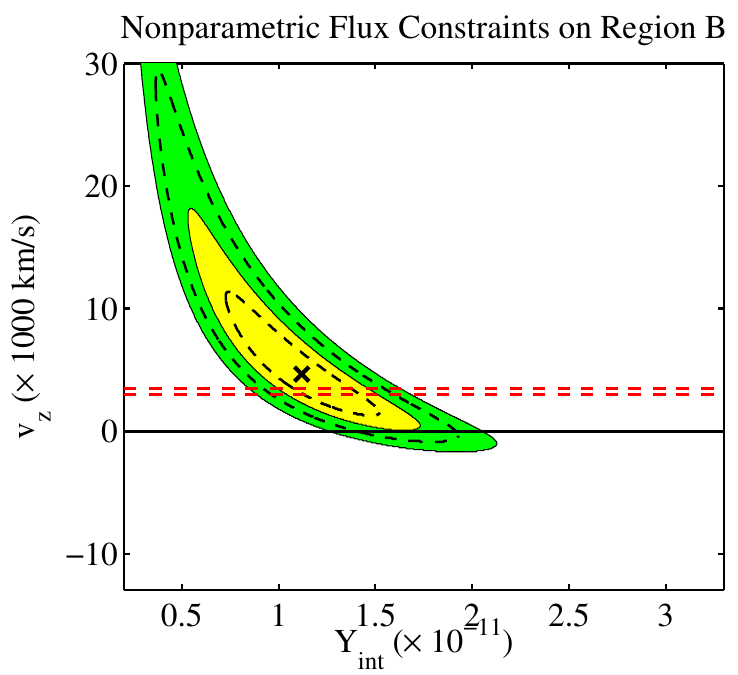}
  } \centerline{
    \includegraphics[width=3.5in]{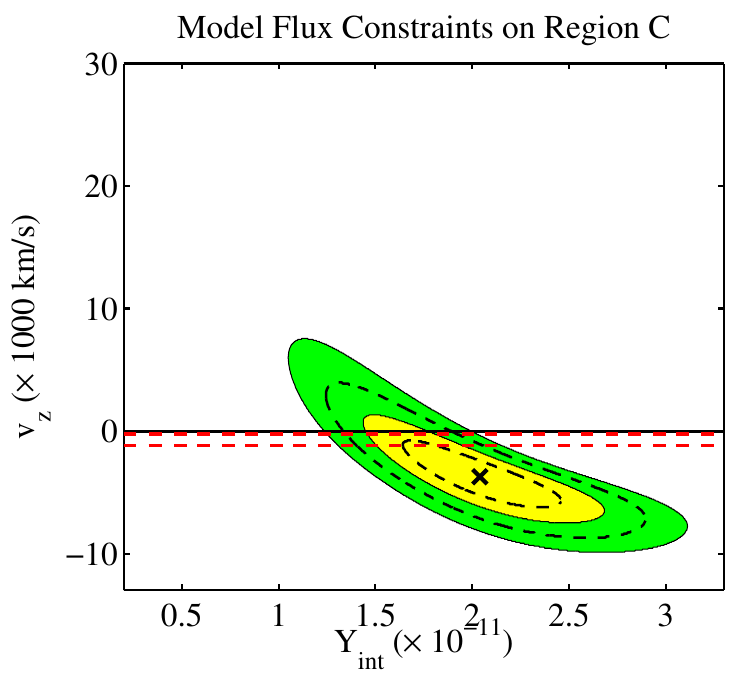}
    \includegraphics[width=3.5in]{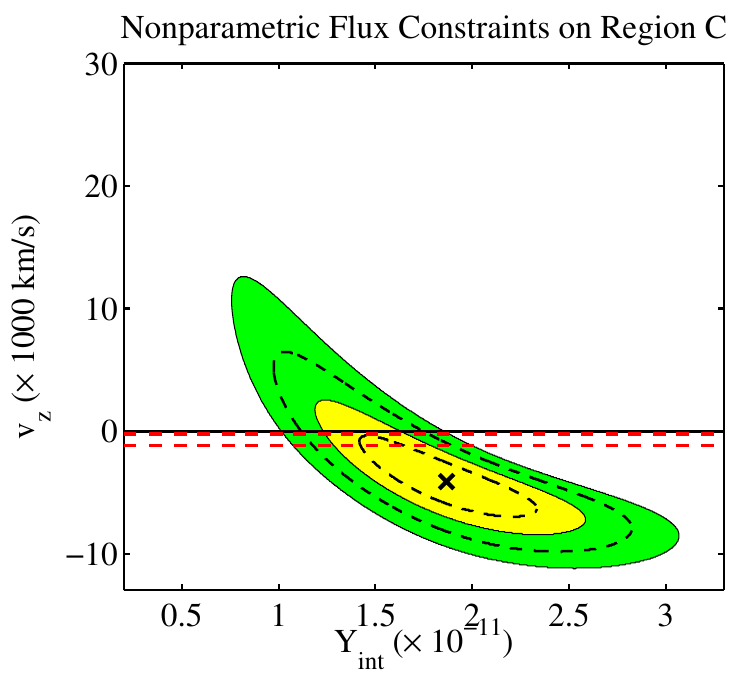}
  }
  \caption{SZE constraints on peculiar velocity and \ysze\ from the Bolocam 140
     and 268~GHz data.  Each subcluster was assumed to be isothermal, and X-ray
     spectroscopic contraints were used to marginalize over temperature.
     Yellow regions contain $\Delta \chi^2 < 2.3$ (68.3\% confidence interval on both 
     parameters), and green regions contain $\Delta \chi^2 < 6.17$ (95.4\% confidence interval on 
     both parameters). 
     Black dashed lines enclose $\Delta \chi^2 < 1$ and $\Delta \chi^2 < 4$, the 
     1 and 2$\sigma$ confidence intervals on the parameters taken individually 
     (i.e. marginalizing over the other parameter). An `X' marks the minimum in $\chi^2$. 
     The horizontal red dashed lines mark 1$\sigma$ constraints from the \cite{ma2009} optical
     spectroscopic velocity measurements.
     {\bf Upper Left:} Constraints on peculiar velocity and \ysze\ using the fluxes measured 
     from the model fits within a 1$^\prime$ diameter region centered on subcluster B.
     {\bf Upper Right:} Same as upper left, but using the nonparametric flux measurements.
     {\bf Lower Left:}  Same as upper left, but for subcluster C.
     {\bf Lower Right:} Same as lower left, but using the nonparametric flux measurements.
  \label{fig:ksz_conts}}
\end{figure*}

\begin{figure*}[t]
  \centerline{
    \includegraphics[width=3.5in]{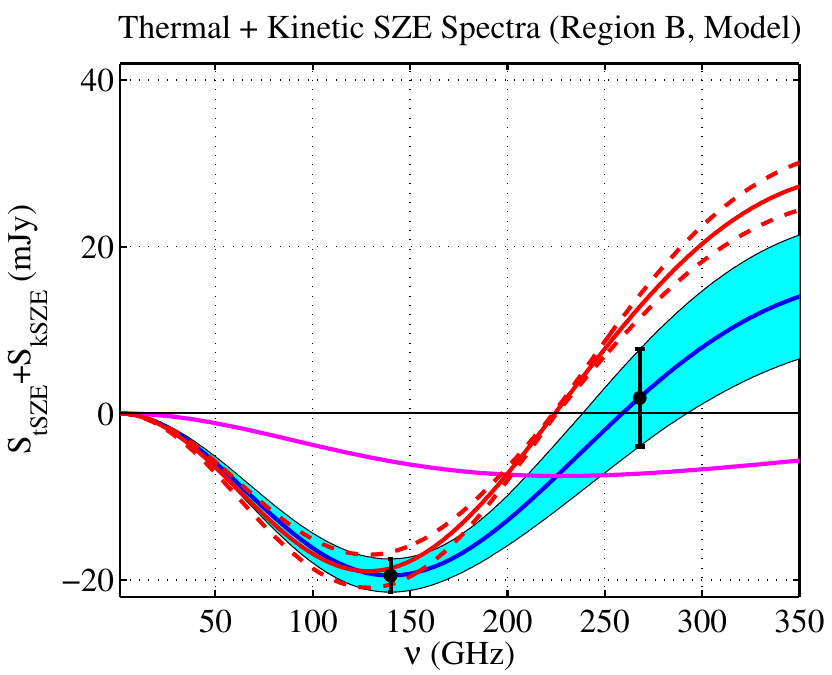}
    \includegraphics[width=3.5in]{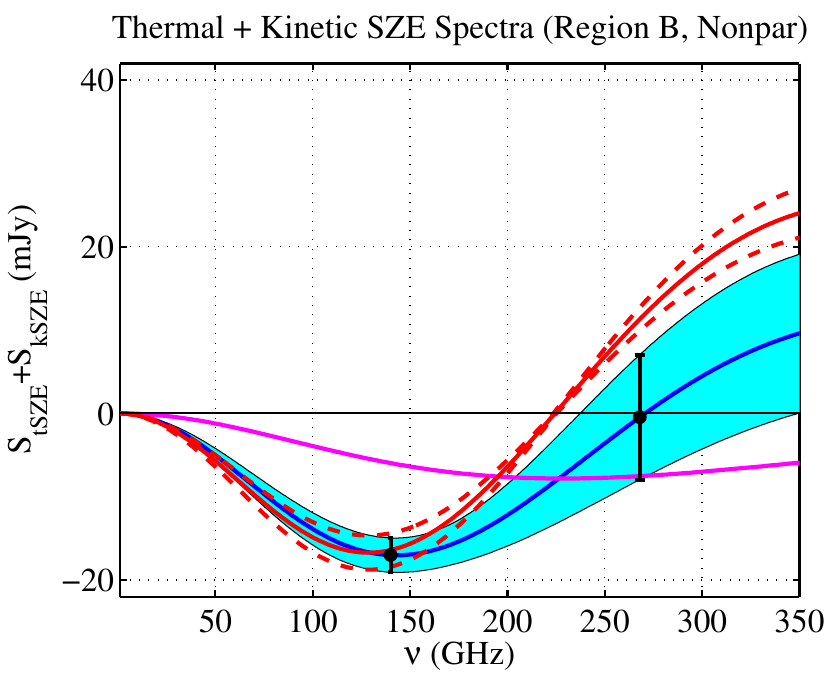}
  }
  \centerline{
    \includegraphics[width=3.5in]{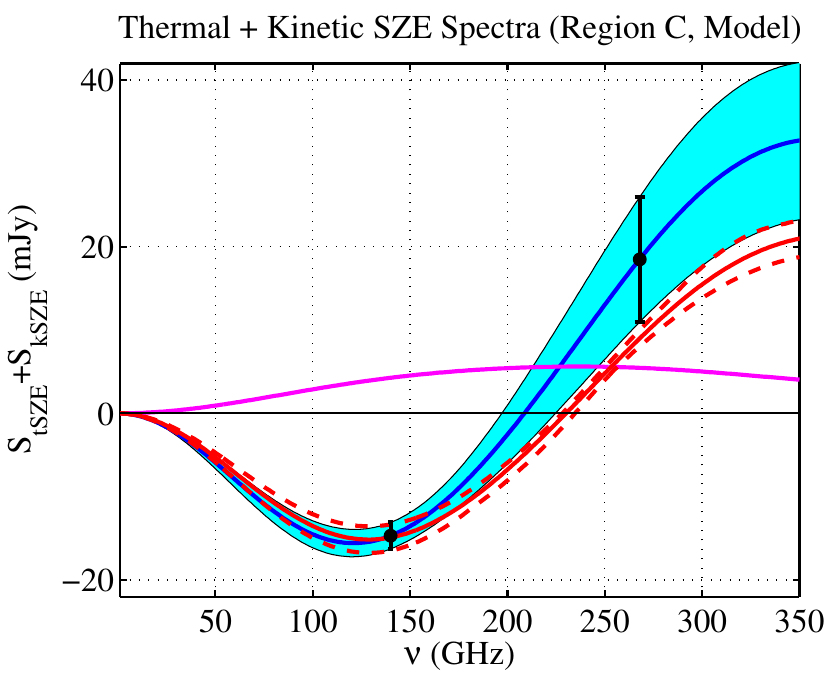}
    \includegraphics[width=3.5in]{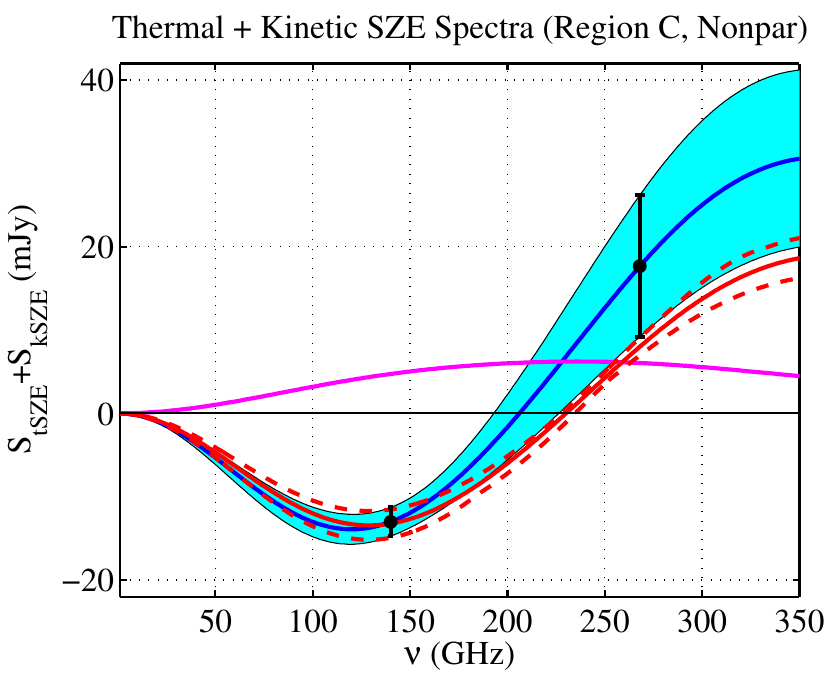}
  }
  \caption{{\bf Upper Left:}  
    Spectral sum of tSZE and kSZE flux densities for
    subcluster B, using the measurements obtained from the model fits to the 
    Bolocam 140 and 268~GHz data (black points with error bars).  
    The best fit combined SZE spectrum is plotted as a solid blue line, with 1$\sigma$ 
    errors displayed as the cyan region.
    SZE spectral fits were obtained through a joint likelihood
    analysis for the Bolocam data including \chandra\ \kB\te\ likelihood
    constraints.
    The kSZE contribution for the best fit velocity in Table~\ref{tbl:ksz} 
    is plotted in magenta (solid line).
    A pure tSZE spectrum to the data is plotted in red (solid line, with dashed 
    lines indicating the 68.3\% confidence interval). 
    Spectra with $v_{z}\leq0$ have a probability of 2.1\% given these data. 
    {\bf Upper Right:} Same as upper left, but using the flux density measured
    directly (nonparametrically) at 140 and 268~GHz from the deconvolved data.
    Spectra with $v_{z}\leq0$ have a probability of 3.4\% given these data. 
    {\bf Lower Left:} Same as upper right, but for subcluster C.
    Spectra with $v_{z}\geq0$ have a probability of 15.7\% given these data. 
    {\bf Lower Right:} Same as upper right, but for subcluster C.
    Spectra with $v_{z}\geq0$ have a probability of 19.3\% given these data. 
    \label{fig:ksz}}
\end{figure*}

Our measurements of the SZE spectra using 140 and 268~GHz Bolocam data are 
sensitive to several possible systematic errors, which we describe here.
Errors due to the relative flux calibration were included in our flux
estimates. Since the regions used in the analysis are 60$^{\prime\prime}$ 
in diameter, and the pointing information in each dataset is accurate to 
better than 5$^{\prime\prime}$, this result cannot be explained by pointing errors.
We also consider possible contamination from dusty, star forming ``submillimeter'' 
galaxies (SMGs) which have been included in our noise model in a statistical sense, 
though bright and/or lensed SMGs could bias our measurements, particularly at higher 
frequencies.
The direction of the submillimeter contamination is key: deviation from a purely thermal 
SZE spectrum from B requires a negative flux density at both 140 and 268~GHz, so 
(positive) contamination from SMGs would cause us to \emph{underestimate} the kSZE 
signal rather than overestimate it. The inferred large, negative proper velocity of 
subcluster C could potentially be affected by a SMG, but is nevertheless 
consistent with the optical velocity to better than 1$\sigma$.   
Importantly, submm {\it Herschel} maps show no indication of contamination at a 
significant level in either region. 
The extended radio emission near C -- too faint at 90 GHz to be seen 
in the much more sensitive MUSTANG observations -- cannot explain the kSZE component 
of subcluster C, nor can compact radio sources, constrained by the MUSTANG observation 
to be $\lesssim 90$~\uJybm\ at 90~GHz.
Finally, temperature substructure due to clumping, merger activity, or the remnant 
core in B could increase the variance in our estimates, but since there is no way 
to constrain the size of this clumping effect with these data we defer to future work.


\section{Conclusions}
\label{sec:conc}

High resolution, multi-wavelength observations of the SZE are now beginning to
offer measurements that are truly complementary to X-ray and optical studies 
of the complicated dynamics in galaxy clusters.  
Here we have presented sensitive, sub-arcminute measurements 
from MUSTANG at 90~GHz and from Bolocam at 140 and 268~GHz.  We compared these 
with lower resolution SZE observations obtained with CARMA/SZA at 31~GHz.
We also compared our SZE observations to the detailed lensing, optical dynamics, 
radio, and our own results using \chandra\ X-ray data to build a two-dimensional
template for modelling the tSZE in this cluster.  

The primary feature in MUSTANG's high-pass-filtered, high-resolution
view of the cluster seems to be associated with the merger activity between two 
subcluster components (C and D in the Figure~\ref{fig:lensing}).  
This feature is also strong in Bolocam's 268~GHz, 31$^{\prime\prime}$-resolution
map of \macsc, and is associated with the hottest gas, which approaches
$\sim 30$~keV in spectral fits to the \chandra\ X-ray data from that region
of the sky.  The feature is bracketed by non-thermal, extended emission from 
the relativistic gas seen in GMRT 610~MHz and VLA 1.4--5~GHz observations, 
providing further supporting evidence for a merger scenario and a multi-phase
intra-cluster medium.
The MUSTANG observation also reveals significant features at subcluster B
which could include contributions from high temperature or density substructures.
MUSTANG's measurements account for $\sim2\%$ of the integrated pressure on 
large scales, as measured in the Bolocam 140~GHz and CARMA/SZA 31~GHz data.

Using the X-ray data to constrain the integrated line-of-sight
pressure, and normalizing it to the large scale SZE observations, we
constructed a two-dimensional template for the tSZE in this cluster.  
The use of the bulk SZE measurements for normalization allows us to convert 
the X-ray pseudo-pressure map into units of Compton-$y$, and yields a value 
for the effective depth of the ICM that is consistent with the cluster's scale
($\sim 1~\rm Mpc$).  While assumptions about the line-of-sight
structure must be made and the model is imperfect, a simple spherical
model for this unvirialized structure is clearly insufficient.
The detailed comparison of these tSZE templates with the SZE data allows 
us to infer the presence of pressure substructure not constrained by X-ray 
observations alone.
Further, we presented for the first time subtraction of compact radio
source contamination from the MUSTANG time-ordered data used in making
our maps.

By subtracting the radio source contamination from the MUSTANG data, and
the tSZE template from both, our observations revealed 
residuals indicating that the X-ray-derived tSZE template provides a poor 
fit that can only qualitatively describe the data.  
For the MUSTANG data, the residuals indicate significant pressure or 
temperature substructure not seen in the X-ray (e.g., out of band hot gas), 
but could also be due to a number of systematics (e.g., clumping, assumption
of a constant line-of-sight depth, temperature substructure, or
filtering and data processing effects).

For the Bolocam data, the residual component at subcluster B after subtraction of the  
template is negative in intensity both below and above the null ($\sim 220$~GHz) 
in the tSZE spectrum.  
The residuals are therefore inconsistent with a purely tSZE component---either 
as an excess or deficit---or any possible compact radio or submillimeter 
source.  
Further, considering the high velocity ($\approx 3200$~km s$^{-1}$) of 
subcluster B, we find this residual to be consistent with the kSZE.  
We note that this measurement is on resolved, subcluster
scales, rather than being due to the proper motion of the cluster as a
whole.  

Using flux densities extracted from our model fits, and marginalizing over the X-ray 
spectroscopic temperature constraints for the region, we found that the high-velocity
subcluster B has a best-fit line-of-sight proper velocity of 
$3600^{+3440}_{-2160}$~km s$^{-1}$.  
This agrees with the optical velocity estimate for the galaxies 
associated with the subcluster.
While our results depend on assumptions about the line-of-sight temperature structure
and the accuracy of the X-ray temperature determination, we find
that the probability $v_z\leq0$ given our data is 2.1\%. 
We also fit the SZE spectrum of the most massive subcluster, C.
For this, we found $-3720^{+2960}_{-2480}$, with a 15.7\% probability a 
$v_z\geq0$ SZE spectrum can describe the region given our data and assumptions.

We also compared our peculiar velocity estimates with spectral fits using 
the fluxes measured nonparametrically from the same 
regions of the maps, and find the probability a $v_z\leq0$ SZE spectrum 
can describe our data for subcluster B is 3.4\%.  
For region C, the probability is 19.3\% that $v_z\geq0$ given
our data. 

This tantalizing result is among the highest significance indications of 
non-zero kSZE from an individual galaxy cluster yet 
\citep[see e.g.,][]{holzapfel1997b,benson2003,mauskopf2012,zemcov2012}.
Even more exciting is that the kSZE spectral component required is on subcluster 
scales.  Clearly, this cluster will be an interesting target for 
both detailed modelling of the dynamics and for future SZE studies.

\acknowledgements

We thank Maxim Markevitch, Simona Giacintucci, Cheng-Jiun Ma, and Dan Coe for 
useful discussions and input.  
We also thank Reinout van Weeren, Adi Zitrin, Annalisa Bonafede, and 
Marceau Limousin for providing radio and lensing maps, and for their input,
which aided tremendously in the interpretation of the cluster astrophysics. 
We thank Marie Rex, Eiichi Egami, and Tim Rawle for their 
support in obtaining the 268~GHz Bolocam observations.

The late night assistance of the GBT operators was much appreciated 
during the observations, as was the overall support from all those
at the GBT.  We also thank Ashley Reichardt for her help in performing
the MUSTANG observations.

We are grateful for CSO administrative support from Kathy Deniston, Barbara 
Wertz, and Diana Bisel, and for Bolocam instrument maintenance and support
from the CSO day crew.

We also thank the many folks who helped in obtaining the CARMA/SZA observations
presented here, particularly Tom Culverhouse, Nikolaus Volgenau, John Carpenter, 
and the many students and postdocs who regularly help run the array.
We are especially grateful to Stephen Muchovej and Erik Leitch for their work 
on the pipeline that allows data from the CARMA/SZA subarray to be easily 
calibrated against Mars. 

Much of the work presented here was supported by National Science Foundation
(NSF) grant AST-1007905. 
Support for T.M.\ was provided by NASA through the Einstein Fellowship Program, 
grant PF0-110077.  
Support for A.Y.\ was provided by the National Radio Astronomy Observatory 
(NRAO) graduate student support program.
Support for P.K.\ was provided by the NASA Postdoctoral Program (NPP).
J.S.\ was supported by NSF/AST-0838261 and NASA/NNX11AB07G. NC was partially
supported by a NASA Graduate Student Research Fellowship.
KU acknowledges support from the Academia Sinica Career Development Award 
and the National Science Council of Taiwan under grant NSC100-2112-M-001-008-MY3. 

The NRAO is a facility of the NSF operated under cooperative agreement by Associated
Universities, Inc. 
The MUSTANG observations presented here were obtained with time on the 
GBT allocated under NRAO proposal IDs AGBT10C017 and AGBT11B001.
The Bolocam data were acquired through observations at the CSO, 
which is operated by the California Institute of 
Technology under cooperative agreement with the NSF (AST-0838261).
A portion of this research was carried out at the Jet Propulsion Laboratory, 
California Institute of Technology, under a contract with the National 
Aeronautics and Space Administration.

We also thank Nina the dog for always standing
by patiently and never questioning the value of this work.


\begin{appendix}
\label{append}

A simple justification for the ``slab approximation,'' which assumes the
  temperature is constant along the line-of-sight, is as
  follows.  The expression for X-ray surface brightness is
\begin{equation}
\label{eq:xray_sb}
\sx = \frac{1}{4\pi (1+z)^3} \! \int \!\! \dene^2 \Lamee(\te,Z) \,d\ell,
\end{equation}
where $\Lamee(\te,Z)$ contains the extra $(1+z)^{-1}$ factor
required by cosmological dimming.
In the absence of any true line-of-sight information about the cluster,
we defer to the standard spherical approximation.
Assuming a standard $\beta$ profile for the density,
\begin{equation}
\dene(r) = n_{\mbox{\tiny e0}}
\left[1+(r/r_c)^2\right]^{-3\beta/2},
\label{eq:beta}
\end{equation}
then surface brightness as a function of sky angle $\theta$ is 
\citep[e.g.,][]{sarazin1988,reese2002,laroque2006}
\begin{equation}
\sx(\theta) = S_{\mbox{\tiny X0}} \left[1+(\theta/\theta_c)^2\right]^{(1-6\beta)/2}.
\end{equation}
For pressure---allowing for the slope $\beta_P$ to be different from 
the density profile's slope $\beta$ (i.e. non-isothermal ICM)---we have 
\begin{equation}
\Pe(r) = P_{\mbox{\tiny e0}}
\left[1+(r/r_c)^2\right]^{-3\beta_P/2}.
\label{eq:betaP}
\end{equation}
Using $y = \sigT/(\mec) \int \Pe d\ell$, Compton-$y$ as a function 
of sky angle $\theta$ is \citep[see e.g.,][]{reese2002,laroque2006}
\begin{equation}
y(\theta) = y_{\mbox{\tiny 0}} \left[1+(\theta/\theta_c)^2\right]^{(1-3\beta_P)/2}.
\end{equation}
Taking $\beta=0.7$ \citep[e.g.,][]{laroque2006} as the average density
slope and $\beta_P=0.86$ for the average pressure slope \citep[see
e.g.,][]{plagge2010,marriage2011b,reese2012}, the ratio of $y(\theta)$
to $\sqrt{\sx(\theta)}$ is now proportional to
$\left[1+(\theta/\theta_c)^2\right]$ raised to a power of
$(1-3\beta_P)/2 - (1-6\beta)/4 = 3(\beta - \beta_P)/2 + 1/4 =
0.01 \approx 0$.  The assumption that the slope is 0 (i.e. the slab approximation
is valid) is within the error bars reported on the average slopes
$\beta$ and $\beta_P$, and prevents the resulting pseudo
Compton-$y$ map from becoming unbounded at the map edges.  

We emphasize that this result does not imply pressure is constant along the line of 
sight, but rather that the average ratio of Compton-$y$ to $\sqrt{\sx}$ 
is approximately constant.  The underlying pressure and density profiles are assumed,
on average, to be described by their respective $\beta$-model parameterizations ($\beta=0.7$
and $\beta_P=0.86$).
If we were to assume $\beta=2/3$ for the density profile instead, this would yield a 
$\beta$-model taper to the power of -0.14.  
Both of these choices of $\beta$, along with $\beta_P=0.86$,
are consistent with a polytropic index of 1.2--1.3, which are in turn 
consistent with values reported in recent studies \citep{bautz2009,capelo2011}.
\end{appendix}

\end{document}